\documentclass{aa}
\usepackage{txfonts}                    

\usepackage{hyperref}        
\usepackage{natbib}
\usepackage{amsmath,amssymb}

\usepackage{graphicx}
\usepackage{ulem}
\usepackage{bm}            
\usepackage{mathtools}

\usepackage{xspace}

\usepackage{makecell}
\usepackage{multirow}

\usepackage{xcolor}

\newcommand{\revise}[1]{{#1}}

\defcitealias{1983SoPh...88..179E}{ER83}
\defcitealias{2002ApJ...577..475R}{RR02}
\defcitealias{2008ApJ...676..717L}{L08}
\defcitealias{2008ApJ...679.1611T}{T08}
\defcitealias{2015A&A...582A.120S}{SL15}
\defcitealias{2023ApJ...943...91W}{paper I}

\newcommand{\Exp}[1]{\ensuremath{{\rm e}^{#1}}}
\newcommand{\mathd}{\ensuremath{{\rm d}}}

\newcommand{\innerp}[2]{\ensuremath{\left<#1 |#2\right>}}

\newcommand{\Alf}{Alfv$\acute{\rm e}$n}

\newcommand{\Schrod}{Schr{\"o}dinger}

\newcommand{\Ac}{\ensuremath{A_{\rm c}}}
\newcommand{\As}{\ensuremath{A_{\rm s}}}

\newcommand{\Ai}{\ensuremath{{\rm Ai}}}
\newcommand{\Bi}{\ensuremath{{\rm Bi}}}

\newcommand{\barkappae}{\ensuremath{\bar{\kappa}_{\rm e}}}
\newcommand{\barki}{\ensuremath{\bar{k}_{\rm i}}}
\newcommand{\barke}{\ensuremath{\bar{k}_{\rm e}}}
\newcommand{\barD}{\ensuremath{\bar{D}}}

\newcommand{\barnui}{\ensuremath{\bar{\nu}_{\rm i}}}
\newcommand{\barnue}{\ensuremath{\bar{\nu}_{\rm e}}}

\newcommand{\kappae}{\ensuremath{\kappa_{\rm e}}}

\newcommand{\kcut}{\ensuremath{k_{\rm cut}}}
\newcommand{\kcutj}{\ensuremath{k_{{\rm cut},j}}}
\newcommand{\ke}{\ensuremath{k_{\rm e}}}
\newcommand{\ki}{\ensuremath{k_{\rm i}}}

\newcommand{\nui}{\ensuremath{\nu_{\rm i}}}
\newcommand{\nue}{\ensuremath{\nu_{\rm e}}}

\newcommand{\omgcrit}{\ensuremath{\omega_{\rm crit}}}

\newcommand{\rhoi}{\ensuremath{\rho_{\rm i}}}
\newcommand{\rhoe}{\ensuremath{\rho_{\rm e}}}

\newcommand{\vA }{\ensuremath{v_{\rm A}}}
\newcommand{\vAe}{\ensuremath{v_{\rm Ae}}}
\newcommand{\vAi}{\ensuremath{v_{\rm Ai}}}

\begin{document}

\title{Temporal evolution of axially standing kink motions in solar coronal slabs: 
       An eigenfunction expansion approach}
\author{          Yuhong Gao            \inst{1}
            \and  Bo Li                 \inst{2,1}
            \and  Mijie Shi             \inst{2} 
            \and  Shaoxia Chen          \inst{2}
            \and  Hui Yu                \inst{2}
       }

\institute{
   Center for Integrated Research on Space Science, Astronomy, and Physics, 
   Institute of Frontier and Interdisciplinary Science, Shandong University, Qingdao 266237, China
   \email{bbl@sdu.edu.cn}
\and 
   Shandong Provincial Key Laboratory of Optical Astronomy and Solar-Terrestrial Environment,
   Institute of Space Sciences, Shandong University, Weihai 264209, China\\
}

\titlerunning{Kink motions in coronal slabs}
\authorrunning{Gao et al.}

\date{Received ......... / Accepted .........}

\abstract
{}
{
We aim to provide more insights into the applicability to solar coronal seismology of the much-studied discrete leaky modes (DLMs) in classic analyses.
}
{Under linear ideal pressureless MHD, we examine two-dimensional (2D) axial fundamental kink motions that arise when localized velocity exciters impact some symmetric slab equilibria. Continuous structuring is allowed for. A 1D initial value problem (IVP) is formulated in conjunction with an eigenvalue problem (EVP) for laterally open systems, with no strict boundary conditions (BCs) at infinity. The IVP is solved by eigenfunction expansion, allowing a clear distinction between the contributions from proper eigenmodes and improper continuum eigenmodes. Example solutions are offered for parameters typical of active region loops.}
{Our solutions show that the system evolves towards 
     long periodicities due to proper eigenmodes
     (of order the axial \Alf\ time), whereas the interference of the improper continuum
     may lead to short periodicities initially (of order the lateral \Alf\ time).
Specializing to the slab axis, we demonstrate that the proper contribution
     strengthens with the density contrast, but may occasionally be stronger for less steep density profiles.
Short periodicities are not guaranteed in the improper contribution, 
     the details of the initial exciter being key. 
When identifiable, these periodicities tend to agree with
     the oscillation frequencies expected for DLMs, despite the differences in the BCs
     between our EVP and classic analyses.
The eigenfunction expansion approach enables all qualitative features 
     to be interpreted as the interplay between the initial exciter and some response function, the latter solely determined by the equilibria.}
{Classic theories for DLMs can find seismological applications, with time-dependent studies 
     offering additional ways for constraining initial exciters.}


\maketitle 

\section{Introduction}
\label{sec_intro}
The past two decades have seen considerable progress in solar coronal seismology
    \citep[e.g., the reviews by][]{2012RSPTA.370.3193D,2020ARA&A..58..441N}.
One key factor has been the increasingly refined theories
    for magnetohydrodynamic (MHD)
    \footnote{\revise{Please see Appendix \ref{sec_App_abbr} for a full list of abbreviations.}} 
    waves in structured media
    \citep[see recent topical collections by][]{2022SSRv..218...13N,2023SoPh..298...40K}.
This can be said even for waves in such canonical equilibria as
    straight cylinders where the physical parameters are structured only radially
    \citep[see the textbooks by][for classic treatments]{2019CUP_Roberts,2019CUP_goedbloed_keppens_poedts}.
Indeed, kink motions have been better
    understood theoretically, facilitating fuller seismological applications
    of, say, the cyclic transverse displacements extensively observed
     in active region (AR) loops 
    \citep[e.g., the dedicated review by][]{2021SSRv..217...73N}.
Likewise, substantial theoretical progress has been made for sausage motions 
    \citep[see the review by][]{2020SSRv..216..136L}, enabling one to better exploit, say, the abundantly measured quasi-periodic pulsations (QPPs) in flares \citep[see e.g.,][for reviews]{2016SoPh..291.3143V,2021SSRv..217...66Z}.

Whatever observable to be used for seismology needs to be offered an unambiguous 
    theoretical understanding, which is not always possible though.
It suffices to regard ``observables''  only as such timescales
    as periods or damping times. 
It also suffices to consider kink motions in a $z$-aligned 
    cylindrical setup
    and consider only those temporal intervals where one may invoke the expectations
    in classic mode analyses. 
By ``classic'' we mean the customary procedure that starts by Fourier-decomposing any
    perturbation as $\exp[-i(\Omega t - kz)]$, where $k$ is the axial wavenumber
    and the possibly complex-valued $\Omega$ is found by seeking nontrivial solutions to some boundary value problem (BVP, see e.g., Chapter 6 in \citealt{2019CUP_Roberts}).
Consider only small $k$ given that AR loops are of primary interest throughout.      
Let ``long'' (``short'') refer to timescales on
    the order of the axial (transverse) \Alf\ time.
The periodicities at large times for kink motions are accepted to be long
    and pertain to the kink frequency $\Re\Omega_{\rm kink}$ (see e.g., Eq.(1)
    in \citealt{1983SoPh...88..179E})
    \footnote{\revise{Here and hereafter, we use $\Re$ and $\Im$ to denote the real and imaginary parts of a complex-valued quantity, respectively.}}, 
    as has been extensively implemented
    \citep[e.g.,][]{1999Sci...285..862N,2001A&A...372L..53N}.
That said, some controversy exists as to whether the accompanying damping is solely
    due to the resonant absorption in the \Alf\ continuum \citep[see][for conceptual clarifications]{2011SSRv..158..289G}. 
Some ``principal fast leaky kink'' mode (PFLK) 
    with $\Re\Omega_{\rm PFLK} \approx \Re\Omega_{\rm kink}$ was additionally found
    in the mode analysis by 
    \citeauthor{1986SoPh..103..277C}~(\citeyear{1986SoPh..103..277C}; see also
    \citealt{1978SoPh...58..165M,1982SoPh...75....3S}). 
Given a non-vanishing $\Im\Omega_{\rm PFLK}$, it was proposed by
    \citet{2003SoPh..217...95C} that PFLK may occasionally play some role in the observed damping. 
However, \citet{2006JPlPh..72..285R} argued otherwise, thereby leading to some extensive
    discussions regarding whether the timescales of PFLKs can make it into the system evolution \citep{2006SoPh..233...79C,2006SoPh..237..119R,2007PhPl...14e2101A}.
Evidently, this controversy may be partially settled by 
    examining kink motions as an initial value problem (IVP). 
\citet{2007SoPh..246..231T} offered such a study, reporting no evidence of PFLKs
    in their 1D solutions found with a finite-element code. 
Instead, temporal signatures were clearly demonstrated for the ``trig'' modes that
    solve the same dispersion relation (DR) that governs PFLKs in \citet{1986SoPh..103..277C}.
Note that some key features of the expected spatial signatures
    can also manifest themselves, as evidenced by a recent 3D numerical study based on the finite-volume approach \citep{2023ApJ...943L..19S}. 
Note further that trig modes are not of concern in the above-mentioned controversy, 
    even though they are subject to some further debate.

Some clarifications on the nomenclature prove necessary. 
By ``classic BVPs'' we refer to those that arise in classic mode analyses; 
    the associated governing equations are ordinary differential equations (ODEs) 
    given that only 1D structuring is of interest.
By ``modes'' we specifically mean the nontrivial solutions to classic BVPs, 
    and a mode is characterized by a pair of ``mode frequency'' ($\Omega$)
    and ``mode function''. 
Let ``spectrum'' refer to the collection of $\Omega$.    
Two reasons are typically responsible for $\Im\Omega$ not to vanish in our context,
    one being the physical relevance of the \Alf\ continuum
    \citep[e.g.,][]{1998PhPl....5.3143G,2011SSRv..158..289G}    
    and the other being that the lateral domain is open.
We focus on the latter.
The boundary condition (BC) at infinity in classic analyses then amounts to
    that no ingoing waves are allowed \citep[e.g.,][]{2007PhPl...14e2101A,2023ApJ...943...91W,2023JPlPh..89e9020G}.
This BC yields two distinct types of mode behavior at large lateral distances, 
    enabling the spectrum to be divided as such. 
One subspectrum comprises the well known ``trapped modes'', 
    the mode functions being evanescent at large distances and 
    the mode frequencies being purely real \citep[$\Im\Omega=0$, e.g.,][]{1982SoPh...76..239E,1983SoPh...88..179E}. 
In contrast, the modes in the other subspectrum are characterized by a non-vanishing
    $\Im\Omega$, and the mode functions are oscillatory.
A rich variety of modes have been reported in this subspectrum, 
    the PFLK and trig modes being examples \citep[e.g., Fig.~1 in][]{2003SoPh..217...95C}.
We concentrate on the trig modes, which feature that the elements in
    the set $\{\Re\Omega\}$ for small axial wavenumbers
    all correspond to short periodicities and are roughly equally spaced
    \citep{1986SoPh..103..277C,2005A&A...441..371T}.
However, we follow \citet{2023ApJ...943...91W}
    to call them ``discrete leaky modes'' (DLMs).
We choose not to use the customary term ``leaky modes'' to avoid confusions; 
    in the literature this term may include PFLKs \citep[e.g.,][]{2003SoPh..217...95C,2006JPlPh..72..285R} or 
    may actually refer to the entire $\Im\Omega\ne 0$ subspectrum
    (\citealt{1986SoPh..103..277C}; also 
    \citealt{1975IGAFS..37....3Z,1982SoPh...75....3S}).

Our study will address kink motions in slab equilibria from an IVP perspective, 
    paying special attention to the relevance of the DLM expectations to our time-dependent solutions.
We stress that the slab geometry, although in less frequent use,
    may be relevant for interpreting the oscillatory behavior measured in, say, 
    post-flare supra-arcades \citep{2005A&A...430L..65V}, streamer stalks \citep[e.g,][]{2010ApJ...714..644C,2011SoPh..272..119F,2020ApJ...893...78D}, and flare current sheets \citep[e.g.,][]{2012A&A...537A..46J,2013A&A...550A...1K,2016ApJ...826...78Y}.
More importantly, key to our study is the wave behavior 
    far from wave-guiding inhomogeneities, meaning that our qualitative findings are applicable to more general geometries. 
That said, DLMs in a slab geometry are indeed the subject of some controversy. 
The classic mode analysis in \citet[][Sect.~5.6]{2019CUP_Roberts} indicates that no kink DLM
    exists that solves the classic BVP when the slab is density-enhanced relative to its ambient corona, in contrast to earlier mode analyses that suggest the opposite
    \citep[e.g.,][]{2005A&A...441..371T,2015ApJ...814...60Y,2018ApJ...855...47C}.
Accepting DLMs as mathematical solutions, 
    one further subtlety is that DLMs may have no physical meaning as argued on energetics grounds by \citet{2023JPlPh..89e9020G}.       
Note that this argument holds regardless of geometries, 
    and hence apparently contradicts
    direct numerical simulations that quantitatively display some expected DLM signatures
    \citep[e.g.,][]{2005A&A...441..371T,2007SoPh..246..231T,2023ApJ...943L..19S}.
Our study is partly motivated by these inconsistencies.

Our essentially analytical approach builds on the standard spectral
    theory of differential operators \citep[e.g.,][]{richtmyer1978principles} or more intuitively on the Fourier-integral-based method \citep[e.g.,][]{1974Whitham}.
Key is that one formulates an eigenvalue problem (EVP) consistent with the IVP
    in both the spatial operator and the BCs.
Given localized initial exciters, what happens at infinity should not affect
    the perturbations at any finite distance. 
Consequently, the BCs at infinity in both our IVP and EVP are only nominal; 
    no definitive requirement is specified. 
Formally speaking, this condition is the only difference between our EVP 
    and the pertinent ``classic BVP''. 
This difference turns out to be crucial, to clarify which we will use
    ``eigen'' in any term that arises in our EVP whenever possible.   
By ``eigenmode'' we refer to a nontrivial solution jointly characterized
     by an eigenfrequency ($\omega$) and an eigenfunction.
By ``eigenspectrum'' we mean the collection of $\omega$. 
We emphasize ``eigen'' because the nontrivial solutions qualify as such, 
    the defining features being that all eigenfrequencies are real-valued
    and the set of eigenfunctions is orthogonal and complete.
Somehow intricate is that the absence of a definitive BC at infinity 
    introduces a continuous eigenfrequency distribution in situations where
    one expects DLMs in classic BVPs.
These details notwithstanding, the wave field at any instant can be decomposed into
    the eigenfunctions, any coefficient depending only on time
    as a simple harmonic oscillator.
The initial conditions (ICs) for the IVP then fully determine all coefficients
    and hence a time-dependent solution. 
We choose to call this approach ``the method of eigenfunction expansion''
    to avoid confusions with the Fourier-decomposition in classic mode analyses.
Although well established, this approach has 
    been adopted in solar contexts only occasionally    (\citealt{2014ApJ...789...48O,2015ApJ...806...56O,2022ApJ...928...33L,2023ApJ...943...91W}; also see e.g., \citealt{1991JPlPh..45..453C,2015ApJ...803...43S,2020ApJ...893..157E} for related studies).
The above-outlined steps were seen as self-evident, 
    but are offered here to clarify matters.  

This manuscript will examine the response of symmetric slab setups
    to localized kink exciters.
We will allow for continuous nonuniformities, but restrict ourselves to strictly 2D motions
    to make the \Alf\ continuum irrelevant \citep[see][for reasons]{2011SSRv..158..289G}. 
We will largely focus on conceptual understandings behind  
    ``How do the periodicities in the $t$-dependent solutions connect to classic mode analysis?'' and ``What role may be played for these periodicities by the equilibrium quantities and initial exciters?''.
Our study is new in the following aspects, relative to the most relevant studies 
    in the literature. 
Firstly, an analytical study dedicated to our purposes with our approach
    is not available.
Similar questions were addressed by \citet{2006JPlPh..72..285R}
    and \citet{2007PhPl...14e2101A} for step profiles.
However, the Laplace transform approach therein is quite involved, and such complications
    as the multi-valued-ness of the Green function may potentially obscure
    the relevance of DLMs.
On the other hand, our approach bears close resemblance to a few available studies,
    which nonetheless also adopted step profiles 
    and were devoted to different contexts \citep[e.g.,][]{2014ApJ...789...48O,2022ApJ...928...33L}.
By addressing continuous profiles, our study will not only demonstrate the generality
    of the eigenfunction expansion approach, but also shed more light on how the profile
    steepness impacts the system evolution. 
Secondly, there exist a considerable number of numerical studies in
    various contexts that partially overlap what to be examined here
    \citep[e.g.,][to name only a few]{2005A&A...441..371T,2007SoPh..246..231T,2012ApJ...761..134N,2016SoPh..291..877G,2020ApJ...893...62L}.
However, the numerical schemes therein are exclusively grid-based, 
    and hence do not explicitly involve such concepts as modes or eigenmodes. 
Our EVP-based approach makes it easier to explore the connection
    between a $t$-dependent solution and modal expectations.
    
This manuscript is structured as follows. 
Section~\ref{sec_prob} presents the steps that eventually lead to a 1D IVP,
    for which the analytical solution is then formulated in Sect.~\ref{sec_modalSol}
    with the eigenfunction expansion method. 
Section~\ref{sec_results} proceeds to present a rather systematic set of example solutions, 
    thereby better visualizing our conceptual understandings. 
Our study is summarized in Sect.~\ref{sec_conc}, where some concluding remarks
    are also offered.

\section{Problem formulation}
\label{sec_prob}

\subsection{Overall description}
\label{sec_sub_2DIBVP}

We adopt pressureless, ideal, MHD throughout, in which the primitive dependents 
   are mass density $\rho$, velocity $\vec{v}$, and magnetic field $\vec{B}$.
Let $(x, y, z)$ denote a Cartesian coordinate system, 
   and let subscript $0$ denote the equilibrium quantities. 
We consider only static equilibria ($\vec{v}_0 = 0$), 
   assuming that the equilibrium magnetic field is uniform and $z$-directed
   ($\vec{B}_0 = B_0 \vec{e}_z$).
The equilibrium density $\rho_0$ is further assumed to 
   be a function of $x$ only and symmetric about $x=0$.
Let $\rhoi$ and $\rhoe$ denote the internal and external densities,
    respectively.
We model density-enhanced coronal slabs
   ($\rhoi > \rhoe$) of half-width $d$ by prescribing
    \begin{eqnarray}
        \rho_0(x)= \rhoi+(\rhoe-\rhoi) f(x), 
                 \label{eq_prof_gen}   
    \end{eqnarray}
    where the function $f(x)$ attains zero at the slab axis ($x=0$)
    but unity outside the slab ($|x|>d$). 
The \Alf\ speed is defined by $v_{\rm A}^2(x) = B_0^2/\mu_0\rho_0(x)$, 
    with $\mu_0$ the magnetic permeability of free space. 
By $\vAi$ ($\vAe$) we denote the internal (external) \Alf\ speed
    evaluated with $\rhoi$ ($\rhoe$). 
We further place two bounding planes at $z=0$ and $z=L$ to
    mimic magnetically closed structures anchored in the dense photosphere.   

We examine axially standing linear kink motions in a strictly two-dimensional (2D) fashion.
Let subscript $1$ denote small-amplitude perturbations.
By 2D we mean $\partial/\partial y=0$, and that we end up with 
   an initial-boundary-value problem (IBVP) involving only 
   $v_{1x}$, $B_{1x}$, and $B_{1z}$.
The equilibria are perturbed in velocity only as implemented by
   the following ICs,
   \begin{subequations}
       \label{eq_2DIBVP_ICs}    
       \begin{align}
           & v_{1x}(x, z, t=0) = u(x) \sin(kz),  \\      
           & B_{1x}(x, z, t=0) = B_{1z}(x, z, t=0) = 0.  
       \end{align} 
   \end{subequations}
Axially standing motions are ensured by prescribing quantized
   axial wavenumbers $k=n\pi/L$ ($n=1, 2, \cdots$),
   and we arbitrarily focus on axial fundamentals ($n=1$).
Kink motions, on the other hand, are ensured by assuming
   the function $u(x)$ in Eq.~\eqref{eq_2DIBVP_ICs} to be even.
It then follows that the 2D IBVP can be examined on 
   the domain where $0\le x <\infty$ and $0 \le z \le L$.
We require that
   $v_{1x}$, $\partial B_{1x}/\partial z$, and $B_{1z}$ vanish  
   at the lower ($z=0$) and upper boundaries ($z=L$).
All dependents are taken to be bounded at $x\to \infty$,
   while the BCs at the slab axis ($x=0$) write
   \begin{equation}
       \label{eq_2DIBVP_BCx0}
         \dfrac{\partial v_{1x}}{\partial x}
       = \dfrac{\partial B_{1x}}{\partial x}
       = B_{1z}
       =0.
   \end{equation}

\subsection{Formulation of the 1D initial value problem}
\label{sec_sub_1DIVP}
Our 2D IBVP simplifies to a simpler 1D version given the ICs and BCs.  
Formally, one may adopt the ansatz
    \begin{subequations}
    \label{eq_Fourier_ansatz}
    \begin{align}
      v_{1x} (x, z,  t) 
    & = \hat{v}  (x, t) \sin(kz), 
    \\
      B_{1x} (x, z, t) 
    &= \hat{B}_x(x, t) \cos(kz), 
    \\
      B_{1z} (x, z, t) 
    &= \hat{B}_z(x, t) \sin(kz), 
    \end{align}
    \end{subequations}
    such that a single equation results for $\hat{v}$,  
    \begin{eqnarray}
         \dfrac{\partial^2 \hat{v}}{\partial t^2}
      &=&
         \vA^2(x)
         \left(
              \frac{\partial^2 \hat{v}}{\partial x^2}
            - k^2 \hat{v}    
         \right).  \label{eq_v2nd_govern}    
   \end{eqnarray}
The ICs write
    \begin{eqnarray}
        \hat{v}(x, t=0) = u(x), \quad  
        \dfrac{\partial \hat{v}}{\partial t} (x, t=0) = 0,
        \label{eq_v2nd_IC}
    \end{eqnarray}
    which follow from Eq.~\eqref{eq_2DIBVP_ICs}.
We note that the 1D problem is in fact an IVP, 
    despite being defined on $0 \le x <\infty$ 
    and subjected to the following BCs,
    \begin{eqnarray}
        \dfrac{\partial\hat{v}}{\partial x}(x=0, t) = 0, \quad
                       \hat{v}(x\to\infty, t) < \infty.
    \label{eq_v2nd_BC}
    \end{eqnarray}   
The BC at the slab axis ($x=0$) follows from parity considerations,
    while the BC at infinity is only nominal.  

\subsection{Parameter specification}
\begin{figure*}
\centering
\includegraphics[width=.95\textwidth]{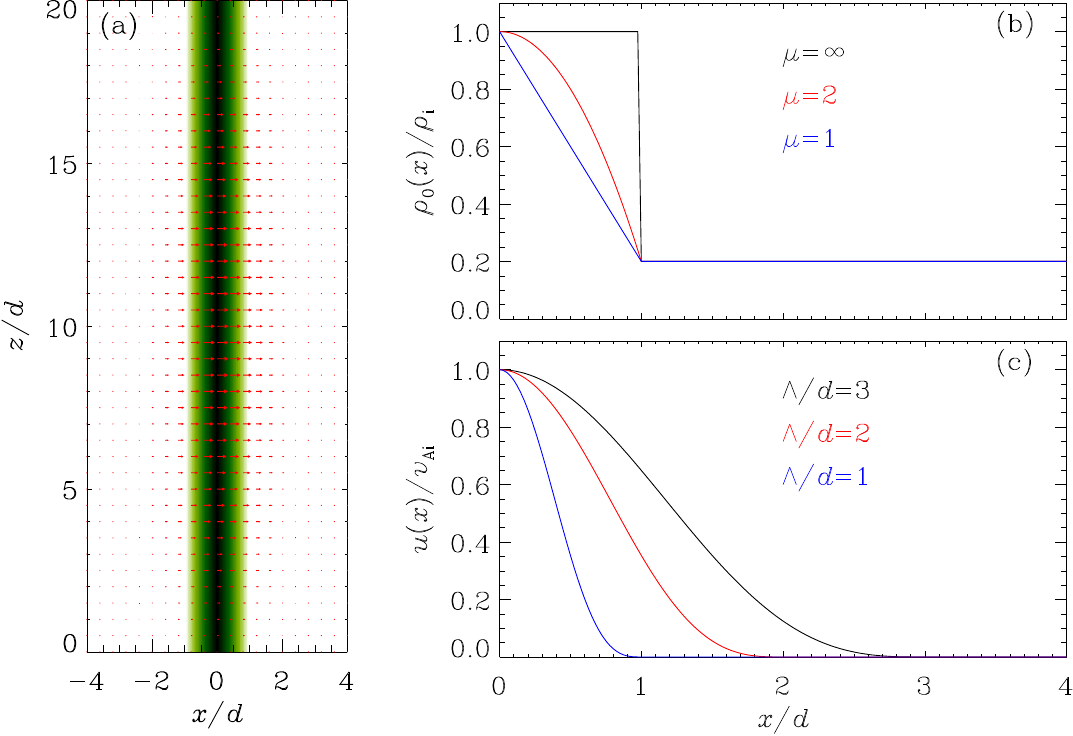}
\caption{
(a) Representation of the two-dimensional (2D) initial boundary value problem
    (IBVP).
The $x-z$ distribution of the equilibrium density is shown by 
    the filled contours, superimposed on which is 
    the initial velocity field (the arrows).
Axial fundamentals are ensured by the $z$-dependence
    of the initial perturbation.
(b) Transverse profiles of the equilibrium density $\rho_0$ 
    as prescribed by Eq.~\eqref{eq_prof_innermu}.
    A number of steepness parameters $\mu$ are considered as labeled, 
        whereas the density contrast is fixed at $\rhoi/\rhoe=5$.
(c) Transverse profiles of the initial perturbation $u(x)$ 
    as prescribed by Eq.~\eqref{eq_u}. 
    Labeled here are a number of values of $\Lambda$, which characterizes the spatial extent
    of the initial exciter.      
 }
\label{fig_Schm} 
\end{figure*}

We proceed to specify the necessary parameters.
To start, we follow \citet{2018ApJ...855...53L} to prescribe
    an ``inner $\mu$'' profile for the density distribution
    by taking $f(x)$ in Eq.~\eqref{eq_prof_gen} to be
    \begin{eqnarray}
    f(x)=\left\{
             \begin{array}{ll}
                  (x/d)^\mu,     	    & 0 \le x \le d, \\[0.2cm]
                  1, 					& x > d,
             \end{array} 
          \right. 
                 \label{eq_prof_innermu}   
\end{eqnarray}
    where $\mu>0$.
Evidently, the density profile becomes steeper with $\mu$, 
    with the much-studied step distribution recovered for $\mu=\infty$.
We further specify the initial exciter in Eq.~\eqref{eq_2DIBVP_ICs} as 
    \footnote{\revise{
    This exciter vanishes for $x > \Lambda$. 
    It follows that $\hat{v}(x,t)$ at a finite $t$ must vanish for $x$ exceeding some $t$-dependent distance, given that perturbations cannot propagate instantaneously. This is actually compatible with the second condition in Eq.~\eqref{eq_v2nd_BC}; both amount to the causality consideration that what happens at infinity should not affect how the fluid behaves at a finite distance.     
    }}
\begin{eqnarray}
\dfrac{u(x)}{\vAi} =
    \left\{
       \begin{array}{ll}
          \cos^3(\pi x/2\Lambda),        & 0 \le x \le \Lambda, 			\\[0.1cm]
          0, 	    			           & x \ge \Lambda.
       \end{array} 
    \right.
 \label{eq_u}
\end{eqnarray}    
This $u(x)$ is rather arbitrarily chosen to be sufficiently smooth,
    with its spatial extent being characterized by the parameter $\Lambda$.
Figure~\ref{fig_Schm}a illustrates our 2D IBVP by displaying the equilibrium density
    $\rho_0$ (the filled contours) and the initial velocity field (arrows).
The $x$-dependencies of $\rho_0$ are shown in Fig.~\ref{fig_Schm}b for
    a number of steepness parameters $\mu$ as labeled, 
    the density contrast being fixed at $\rhoi/\rhoe=5$.  
Likewise, the function $u(x)$ is displayed in Fig.~\ref{fig_Schm}c 
    for an arbitrarily chosen set of $\Lambda$.

Technical details aside, linear kink motions 
    are fully dictated by two sets of parameters.
The dimensional set is taken to be $\{\rhoi, d, \vAi\}$, 
   which serves merely as normalizing constants.
The behavior of linear motions then hinges on the dimensionless set, 
    \begin{equation}
    \label{eq_DimAnalysis}
        \{\rhoi/\rhoe, \mu; kd=\pi d/L; \Lambda/d \}.
    \end{equation}
The effects of $\mu$ and $\Lambda/d$ are of primary interest,
    and we also adjust the density contrast $\rhoi/\rhoe$ when necessary. 
With AR loops in mind, we fix the axial wavenumber at $kd=\pi/20$.
For axial fundamentals, this corresponds to an axial length $L=20d$,  
    which lies somewhat toward the lower end of the range
    for AR loops typically imaged in the EUV
    \citep[e.g.,][]{2004ApJ...600..458A,2007ApJ...662L.119S}. 
The density contrast $\rhoi/\rhoe$ is taken to be in the range $[2, 10]$,
    which is representative of AR loops as well
    \citep[e.g.,][and references therein]{2004ApJ...600..458A}.

\section{EVP-based solutions: formalism}
\label{sec_modalSol}
This section solves the 1D IVP formulated 
   in Sect.~\ref{sec_sub_1DIVP} by the method of eigenfunction expansion.

\subsection{EVP-based solutions for generic transverse structuring}
\label{sec_sub_SolOverall}
As described in the Introduction, our approach is based on the general 
    spectral theory of differential operators.
In principle, there is no need for the initial exciter $u(x)$
    to follow Eq.~\eqref{eq_u} or for the equilibrium density $\rho_0(x)$ to follow Eq.~\eqref{eq_prof_innermu}.
This subsection specializes to the ``inner $\mu$''
    prescription of $\rho_0(x)$ for definitiveness;
    nonetheless, the steepness parameter $\mu$ is allowed to be arbitrary. 
The corresponding formulations generalize those in Sect.~3.1 
    of \citet{2023ApJ...943...91W} where we handled $t$-dependent sausage motions in coronal slabs with step profiles.

Key to our approach is the EVP 
    that pertains to the spatial operator $\mathcal{L}$ of the 1D IVP
    (see Eq.~\eqref{eq_v2nd_govern}).
Specifically, nontrivial solutions are sought for the equation 
    \begin{equation}
        \mathcal{L}\breve{v}
        \coloneqq
        -\vA^2(x)
             \left( 
    	          \dfrac{\mathd^2}{\mathd x^2}\breve{v} 
     			      - k^2 \breve{v}
              \right)
    = \omega^2 \breve{v}, 
    \label{eq_modal_EVP_govE}
    \end{equation}
   defined on $[0, \infty)$ and subject to the BCs (see Eq.~\eqref{eq_v2nd_BC})
    \begin{equation}
       \mathd\breve{v}/\mathd x (x=0) = 0, \quad
       \breve{v}(x\to\infty) < \infty.
    \label{eq_modal_EVP_BC}
    \end{equation} 
Note that $k>0$ serves as an arbitrary parameter here.

One recognizes that the operator $\mathcal{L}$ is \revise{self-adjoint} under the definition
    of the scalar product 
    \begin{eqnarray}
    \label{eq_def_innerprod}
        \innerp{U}{V}
    \coloneqq    
        \int_{0}^{\infty} U^*(x)V(x) \rho_0(x) \mathd x,
    \end{eqnarray}    
    where the asterisk represents complex conjugate.  
Some properties then readily follow from general theory 
    \revise{(see, in particular, \citealt{2007PhPl...14e2101A} for in-depth discussions on the self-adjoint-ness)}.
\begin{itemize}
\item All eigenvalues $\omega^2$ are positive, and we see the eigenfrequency
       $\omega$ as positive without loss of generality.
\item All eigenfunctions can be made and are seen as real-valued.
\item The eigenspectrum, namely the collection of $\{\omega\}$, comprises
      a point subspectrum and a continuum subspectrum.
\item The eigenmodes in the point subspectrum are jointly characterized
          by some eigenfrequency $\omega_j$ and eigenfunction $\breve{v}_j$,
          with $j=1,2,\cdots$ being the mode label. 
      The eigenfrequency $\omega_j$ is dictated by some DR, 
          satisfying $k \vAi< \omega_j  < k \vAe \coloneqq \omgcrit$. 
      The eigenfunction $\breve{v}_j$ is evanescent for $x>d$, 
          making the modes ``proper'' in 
          \revise{that $\breve{v}_j$ is square integrable (i.e., $\innerp{\breve{v}_j}{\breve{v}_j}$ converges in classic sense)}. 
\item Proper eigenmodes are subject to cutoff wavenumbers 
          $0<k_{{\rm cut}, 2}<k_{{\rm cut}, 3}<\cdots$
          with the exception of the lowest mode $j=1$.
      By cutoff we mean that only one proper eigenmode arises 
          when $0< k< k_{{\rm cut}, 2}$, while a total of $J$ proper eigenmodes are allowed when $k_{{\rm cut}, J} < k <k_{{\rm cut}, J+1}$ for $J\ge 2$.   
\item The eigenmodes in the continuum subspectrum are jointly characterized
          by some eigenfrequency $\omega$ and eigenfunction $\breve{v}_\omega$.
      The eigenfrequency $\omega$ continuously spans
          $(\omgcrit, \infty)$, meaning the irrelevance of
          the concept of DRs.
      The eigenfunction $\breve{v}_\omega$ is oscillatory for $x>d$, 
          making the modes ``improper'' in that 
          \revise{$\breve{v}_\omega$ is not square integrable (i.e., $\innerp{\breve{v}_\omega}{\breve{v}_\omega}$ is not defined in classic sense)}.
\item  The eigensolutions in the proper set and the improper continuum 
           are complete, satisfying the orthogonality condition
        \begin{subequations}
        \label{eq_def_Ortho}
        \begin{align} 
        &     \innerp{\breve{v}_j}{\breve{v}_{j'}} 
            = \innerp{\breve{v}_j}{\breve{v}_j} \delta_{j,j'},  \\
        &     \innerp{\breve{v}_j}{\breve{v}_\omega} 
            = 0,  \\
        &     \innerp{\breve{v}_\omega}{\breve{v}_{\omega'}} 
            = q(\omega) \delta(\omega-\omega'),
        \end{align}
        \end{subequations}
             where $\delta_{j,j'}$ is the Kronecker delta.  
\end{itemize}
   
The EVP-based solution to our 1D IVP eventually writes
    \begin{equation}
    	\label{eq_modalSol}      
    \begin{split}
    &    \hat{v}(x, t)
       = \sum\limits_{j = 1}^{J} c_j     
              \breve{v}_j(x)      \cos(\omega_j t) 
       + \int_{k \vAe}^{\infty}  S_\omega 
    	      \breve{v}_\omega(x) \cos(\omega   t) \mathd\omega,
    \\
    &   0\le x < \infty, \quad 0\le t < \infty.	      
    \end{split}	
    \end{equation}       
We have made it explicit that 
    the solution \eqref{eq_modalSol} is valid for arbitrary $x$ and $t$. 
The summation on the right-hand side (RHS) collects all (up to $J$) proper eigenmodes 
    that may arise for a given $k$, 
    while the integration accounts for the contribution from improper eigenmodes. 
The coefficients, $c_j$ and $S_\omega$, are given by
    \begin{equation}
    \label{eq_Sol_DefcjSomg}
    c_j = \dfrac{\innerp{u          }{\breve{v}_j}}
      		 {\innerp{\breve{v}_j}{\breve{v}_j}}, \quad 
    S_\omega =                	     
          \dfrac{\innerp{u          }{\breve{v}_\omega}}
    	    	{q(\omega)}.
    \end{equation}
We will refer to $S_\omega \breve{v}_\omega(x)$ as
    some ``local spectral density''.

Some general expectations can be made with Eq.~\eqref{eq_modalSol}
    regarding the possible relevance of DLMs.
We recall that the concept of DRs applies only to
    proper eigensolutions to our EVP.
However, the proper and improper eigenmodes are not distinguished beforehand. 
Rather, this distinction arises naturally to comply with the nominal BC at infinity,
    where actually no restriction is imposed. 
We recall further that the BC at infinity in ``classic BVPs'' is that no ingoing waves
    are allowed therein.
Our Appendix~\ref{sec_App_DLM} collects some necessary examinations on classic BVPs, 
    and the concept of DRs is shown to always apply. 
Two types of nontrivial solutions arise, with the trapped modes being identical to
    our proper eigenmodes.
DLMs, on the other hand, are demonstrated to exist 
    for all the examined density profiles, thereby corroborating and generalizing
    available findings by, say, \citet{2023JPlPh..89e9020G}.
Now consider the oscillation frequencies of DLMs $\{\Re\Omega^{\rm DLM}\}$.
Given Eq.~\eqref{eq_modalSol}, there is clearly no guarantee 
    for $\{\Re\Omega^{\rm DLM}\}$ to stand out in a $t$-dependent solution;
    DLMs pertain to the classic BVP rather than our EVP.
In this regard, we agree with \citet{2023JPlPh..89e9020G} in saying that 
    DLMs do not have any physical meaning per se.  
However, we argue that the theories on DLMs may still prove seismologically relevant
    if such timescales as $\{\Re\Omega^{\rm DLM}\}$ can be identified
    in some $t$-dependent solutions.
Evidently, for DLMs to be relevant, the set $\{\Re\Omega^{\rm DLM}\}$
    needs to play some special role for the frequency-dependence of $S_\omega$ in 
    Eq.~\eqref{eq_modalSol}. 
Equally evident is that the details of the initial exciter must be crucial
    in determining whether this relevance materializes (see Eq.~\eqref{eq_Sol_DefcjSomg}).

Some further remarks are now necessary, 
    given their immediate relevance to the ``inner $\mu$'' prescription
    in Eq.~\eqref{eq_prof_innermu}. 
We start with the following definitions,
    \begin{subequations}
     \label{eq_def_kike}
    \begin{align}
    &  \ki^2 \coloneqq \dfrac{\omega^2 - k^2\vAi^2}{\vAi^2} 
                     = \dfrac{\omega^2}{\vAi^2}-k^2, \\
    &  \ke^2 \coloneqq \dfrac{\omega^2 - k^2\vAe^2}{\vAe^2} 
                     = \dfrac{\omega^2}{\vAe^2}-k^2, \\
    &  \kappae^2 \coloneqq -\ke^2 
                         = k^2-\dfrac{\omega^2}{\vAe^2}, \\    
    &  D \coloneqq \ki^2+\kappae^2 
                 = \dfrac{\omega^2}{\vAi^2}-\dfrac{\omega^2}{\vAe^2}.  
    \end{align}
    \end{subequations}
Note that $\ki^2$ and $D$ are positive definite, 
     and we see $\ki$ as positive.
Some intricacy arises for $\ke^2$.  
We choose to work with $\ke^2$ and see $\ke$ as positive 
     when handling improper eigenmodes ($\ke^2>0$).
For proper eigenmodes ($\ke^2 < 0$), however,
     we always opt for $\kappae^2$ and see $\kappae$ as positive.
Regardless, the following dimensionless quantities are further defined,
    \begin{subequations}
    \label{eq_def_barkike}
    \begin{align}
    &   \barki        \coloneqq \ki d, \quad  
        \barke        \coloneqq \ke d,     \\
    &   \barkappae    \coloneqq \kappae d, \quad 
        \barD         \coloneqq D d^2,  
    \end{align}
    \end{subequations}
    which will be used only when absolutely necessary.
    
We proceed to reformulate 
    Eq.~\eqref{eq_modal_EVP_govE} into a \Schrod\ form,
\begin{equation}
    \dfrac{\mathd^2 \breve{v}}{\mathd x^2}
   +\left[\dfrac{\omega^2}{\vA^2(x)}-k^2\right]
    \breve{v}
   =0. 
   \label{eq_evp_Schrod}
\end{equation}
Some general results can be obtained 
    for the uniform exterior
    ($\vA(x) =\vAe$ for $x>d$).  
The external solution always writes 
\begin{equation}
    \label{eq_extsol_proper}
    \breve{v}_j(x) \propto \Exp{-\kappae x}
\end{equation}
    for proper eigenmodes, and is always expressible as
    \begin{equation}
    \label{eq_extsol_generic}
        \breve{v}_\omega (x)
     =  \vAi 
        \left[\Ac \cos(\ke x) + \As \sin(\ke x) \right]
    \end{equation}  
    for improper eigenmodes.
Here $\Ac$ and $\As$ are some constants of integration.  
The function $q(\omega)$ in Eq.~\eqref{eq_Sol_DefcjSomg} evaluates to 
    \begin{equation}
    \label{eq_qomg_generic}
      q(\omega) 
    = (\rhoe \vAi^2) 
      \dfrac{\ke \vAe^2}{\omega} 
      \dfrac{\pi(\Ac^2+\As^2)}{2}.
    \end{equation} 
Equation~\eqref{eq_qomg_generic} was first given in our previous sausage study
    \citep[][Eq.~(34)]{2023ApJ...943...91W}, the derivation benefiting substantially
    from Appendix B of \citet{2015ApJ...806...56O}.
Note that sausage and kink eigensolutions in our setup 
    formally differ only in the interior ($x<d$).
The reason for Eq.~\eqref{eq_qomg_generic} to involve only the external solution
    is that the integral over the interior 
    is regular and hence does not contribute (see Eqs.~\eqref{eq_def_innerprod} and \eqref{eq_def_Ortho}).
Note also that Eqs.~\eqref{eq_extsol_generic} and \eqref{eq_qomg_generic}
    apply to all steepness parameters ($\mu$), whose effects show up only indirectly
    via $\Ac$ and $\As$.     
    
Now consider the interior ($x<d$). 
We examine only a representative set of $\mu$, for which
    compact analytical solutions exist for Eq.~\eqref{eq_evp_Schrod}. 
Regardless, the following aspects always hold. 
\begin{itemize}
    \item An internal eigenfunction is found by solving Eq.~\eqref{eq_evp_Schrod}
          in conjunction with the BC at $x=0$ (see Eq.~\eqref{eq_modal_EVP_BC}). 
    \item The magnitude of an eigenfunction is irrelevant in principle.  
    In practice, we nonetheless scale any internal improper eigenfunction
         $\breve{v}_\omega$ in such a way that it remains regular for vanishingly small $\ke$ as happens
    when $\omega \to \omgcrit=k\vAe$.     
    
    \item The internal solution is connected to the external one by demanding the continuity of both $\breve{v}$ and $\mathd \breve{v}/\mathd x$ 
    at $x=d$. 

    \item We always start with improper eigenmodes. 
    Proper eigenmodes will be described afterwards.
    The expression for a proper eigenfunction $\breve{v}_j$ is written to ensure
        the continuity of $\breve{v}_j$ itself, and the continuity of 
        $\mathd \breve{v}_j/\mathd x$ then yields a DR. 
    Expressions for $\innerp{\breve{v}_j}{\breve{v}_j}$ will be 
        offered when available. 
    We will also analytically examine cutoff axial wavenumbers $\kcutj$, 
         but refrain from fully analyzing the DRs.
\end{itemize}

\subsection{The case with $\mu=\infty$}
\label{sec_sub_SolStep}
This subsection examines the much-studied step profile ($\mu=\infty$).
The internal solution to Eq.~\eqref{eq_evp_Schrod} is simply
     $\propto \cos(\ki x)$
     given that the interior is uniform.
An improper eigenfunction $\breve{v}_\omega (x)$ writes    
\begin{eqnarray}
    \dfrac{\breve{v}_\omega (x)}{\vAi}
  =  
    \left\{
       \begin{array}{ll}
           \dfrac{\ke}{k_i}  \cos(\ki x),   
		        & \quad 0 \le x \le d, 			\\[0.3cm]
       
           \Ac \cos(\ke x) + \As \sin(\ke x),   
				& \quad x > d,
     \end{array} 
    \right.
 \label{eq_step_vImproper}
\end{eqnarray}        
   where 
\begin{equation}
\label{eq_step_ACAS}           
\begin{split} 
& \Ac =  \dfrac{\ke}{\ki} \cos(\ki d) \cos(\ke d) 
                        + \sin(\ki d) \sin(\ke d), \\
& \As =  \dfrac{\ke}{\ki} \cos(\ki d) \sin(\ke d) 
                        - \sin(\ki d) \cos(\ke d).
\end{split}
\end{equation} 

Now consider proper eigenmodes. 
An eigenfunction writes 
\begin{eqnarray}
  \dfrac{\breve{v}_j (x)}{\vAi}=  
  \left\{
    \begin{array}{ll}
       \Exp{-\kappae d} \cos(\ki x),   
		        & \quad 0 \le x \le d, 			\\[0.3cm]
       \cos(\ki d) \Exp{-\kappae x},   
				& \quad x > d.
    \end{array} 
 \right.
 \label{eq_step_vproper}
\end{eqnarray}
A rather concise expression is available for $\innerp{\breve{v}_j}{\breve{v}_j}$, 
    reading  
    \begin{equation}
    \label{eq_step_innerPproper}
      \dfrac{\innerp{\breve{v}_j}{\breve{v}_j}}{\rhoi \vAi^2 d}
    = \dfrac{\Exp{-2 \kappae d}}{2} 
      \left[
            1
          + \dfrac{\sin(2\ki d)}{2\ki d}
          + \dfrac{\cos^2(\ki d)}{\kappae d}\dfrac{\rhoe}{\rhoi}      
      \right].    
    \end{equation} 
The DR governing the eigenfrequency $\omega_{j}$ writes
    \begin{eqnarray}
        \ki\tan(\ki d) = \kappae.
       \label{eq_step_properDR}
    \end{eqnarray} 
Cutoff wavenumbers ($\kcutj$) arise when the axial phase speed
   $\omega/k$ approaches $\vAe$ in response to a varying $k$.
Now that $\kappae \to 0^+$, one recognizes from Eq.~\eqref{eq_step_properDR} that
    \begin{equation}
       \kcutj d= \dfrac{(j-1)\pi}{\sqrt{\rhoi/\rhoe-1}},
               \quad (j=2,3,\cdots).
       \label{eq_step_kcut}
    \end{equation}
The expressions for both the DR (Eq.~\eqref{eq_step_properDR})
    and cutoff wavenumbers (Eq.~\eqref{eq_step_kcut}) 
    are standard textbook material
    \citep[e.g.,][Chapter 5]{2019CUP_Roberts}, albeit largely
    in the context of classic BVPs. 
    
\subsection{The case with $\mu=2$}
\label{sec_sub_Sol_mu02}
This subsection addresses the case where $\mu=2$,
    for which purpose the following definitions are necessary,
    \begin{subequations}
    \label{eq_mu02_def_pX}
    \begin{align}
        & p \coloneqq \sqrt{\barD} 
                    = \dfrac{\omega d}{\vAi}\sqrt{1-\dfrac{\rhoe}{\rhoi}}, \\
        & \alpha \coloneqq \dfrac{1}{4}-\dfrac{{\barki}^2}{4p}
                         = \dfrac{1}{4}-\dfrac{(\omega d/\vAi)^2-(kd)^2}{4p}, \\[0.2cm]
        & X \coloneqq p (x/d)^2.                 
    \end{align}
    \end{subequations}
The internal solution always writes $\Exp{-X/2} M(\alpha, 1/2, X)$ 
    to respect the BC at $x=0$, with $M(\cdot, \cdot, \cdot)$ 
    being Kummer's $M$ function
    \citep[e.g.,][Chapter 13]{NIST:DLMF}.      
An improper eigenfunction writes
    \begin{eqnarray}
      \dfrac{\breve{v}_\omega (x)}{\vAi}
    = \left\{
        \begin{array}{ll}
             \barke \Exp{-X/2}{M(\alpha,1/2,X)} ,   
    		        & \quad 0 \le x \le d, 			\\[0.3cm]
             \Ac \cos(\ke x) + \As \sin(\ke x),   
    				& \quad x > d,
        \end{array} 
     \right.
    \end{eqnarray}
    where
    \begin{subequations}
    \label{eq_mu02_ACAS}
    \begin{align} 
        & \Ac = \barke \Exp{-X_1/2} M(\alpha,1/2,X_1)\cos(\ke d) 
                    -Q\sin(\ke d), \\ 
        & \As = \barke \Exp{-X_1/2} M(\alpha,1/2,X_1)\sin(\ke d)   
                    +Q\cos(\ke d). 
    \end{align}
    \end{subequations}
Furthermore, $X_1=p$ is the value of $X$ at $x=d$, 
    and the symbol $Q$ is defined by
    \begin{equation}
        Q = - p         \Exp{-X_1/2} M(\alpha,   1/2, X_1)
            + 4 p\alpha \Exp{-X_1/2} M(\alpha+1, 3/2, X_1).
    \end{equation} 
Here $M(\alpha+1, 3/2, X_1)$ appears because of the identify
\begin{equation}
     \dfrac{\mathd}{\mathd \mathfrak{z}} 
                  M(a,   b,   \mathfrak{z})
   = \dfrac{a}{b} M(a+1, b+1, \mathfrak{z}).
\end{equation}

Now consider proper eigenmodes. 
An eigenfunction writes
    \begin{eqnarray}
      \dfrac{\breve{v}_j (x)}{\vAi}
    =  
      \left\{
        \begin{array}{ll}
              \Exp{-X/2}   M(\alpha, 1/2,X)   \Exp{-\kappae d},   
    		        & \quad 0 \le x \le d, 			\\[0.3cm]
              \Exp{-X_1/2} M(\alpha, 1/2,X_1) \Exp{-\kappae x},   
    				& \quad x > d.
        \end{array} 
     \right.
     \label{eq_mu02_vproper}
    \end{eqnarray}
A compact expression is not available for
    $\innerp{\breve{v}_j}{\breve{v}_j}$.
However, a DR may be readily derived, reading
\begin{equation}
      -\barkappae 
    = -p + 4p\alpha \dfrac{M(\alpha+1, 3/2, p)}{M(\alpha, 1/2, p)}.
    \label{eq_mu02_DR}
\end{equation}
The DR for trapped sausage modes in the same equilibrium was first given
     by \citet{1988A&A...192..343E}.
However, the DR for kink motions (Eq.~\eqref{eq_mu02_DR}), is new as far as we are aware. 

We proceed to derive the approximate expressions for cutoff wavenumbers $\kcutj$.
Interestingly, this derivation is connected to the behavior at large axial wavenumbers
    ($kd\to\infty$), in which case $\omega/k\to \vAi$.
One readily deduces that $\barkappae\to \infty$, $p\to \infty$,
    and $\barkappae/p\to 1^-$, given their definitions 
    (see Eqs.~\eqref{eq_def_kike}, \eqref{eq_def_barkike}, and \eqref{eq_mu02_def_pX}).
The DR \eqref{eq_mu02_DR} then dictates that 
    \begin{equation}
    \begin{split}
        E &\coloneqq 
            \alpha \dfrac{M(\alpha+1, 3/2, p)}{M(\alpha, 1/2, p)} \\
          & = \alpha 
            \dfrac{1+\dfrac{\alpha+1}{3/2} p
                    +\dfrac{(\alpha+1)(\alpha+2)}{(3/2)(3/2+1) 2!} p^2
                    +\cdots}
                  {1+\dfrac{\alpha}{1/2} p
                    +\dfrac{(\alpha)(\alpha+1)}{(1/2)(1/2+1) 2!} p^2
                    +\cdots} \\
           & \to 0,
    \end{split}
        \label{eq_mu02_E}
    \end{equation}
    where the second equal sign follows from
    the definition of Kummer's $M$ function.
Evidently, Eq.~\eqref{eq_mu02_E} is guaranteed provided 
    $\alpha \to 0, -1, -2, \cdots$.
One may then formally write 
    \begin{equation}
         \alpha_j \to 1-j,    \quad \mbox{for } kd\to\infty,
         \label{eq_mu02_alphaj}
    \end{equation}
    where $j=1, 2, \cdots$ is recalled to be the mode label.
It turns out that Eq.~\eqref{eq_mu02_alphaj}, while derived for $kd\to\infty$,
    holds approximately even for relatively small values of $kd$.
Note that $\alpha$ evaluates to $(1-\kcut \sqrt{\rhoi/\rhoe-1})/4$ at cutoffs
    given that $\omega/k = \vAe$ (see Eq.~\eqref{eq_mu02_def_pX}).
Approximating this $\alpha$ with Eq.~\eqref{eq_mu02_alphaj} then yields that
    \begin{equation}
               \kcutj d 
       \approx \dfrac{4j-3}{\sqrt{\rhoi/\rhoe-1}}, \quad 
            (j=2, 3, \cdots),
       \label{eq_mu02_kcut}  
    \end{equation}
    where the absence of $k_{{\rm cut}, 1}$ is accounted for.
This approximate expression is increasingly accurate when $j$ increases,
    overestimating the exact value by $16.6\%$, $8.3\%$, and $5.5\%$
    for $j=2, 3$ and $4$, respectively.

\subsection{The case with $\mu=1$}
\label{sec_sub_Sol_mu01}
This subsection addresses the case where $\mu=1$, for which purpose the following
    definition is necessary
    \begin{equation}
        X \coloneqq 
          \dfrac{-\barki^2+\barD(x/d)}{\barD^{2/3}}.
    \end{equation}
The internal solution is then a linear combination
    of Airy's functions $\Ai$ and $\Bi$, reading
    \begin{equation}
          \breve{v}(x) 
     \propto 
          \dfrac{\Ai(X)}{\Ai'(X_0)}
         -\dfrac{\Bi(X)}{\Bi'(X_0)}
    \end{equation}
    with $\Ai'$ and $\Bi'$ being Airy's prime functions such that
    the BC at $x=0$ is respected
    \citep[e.g.,][Chapter 9]{NIST:DLMF}. 
Furthermore, 
    \begin{equation}
        X_0 = \dfrac{-\barki^2}{\barD^{2/3}}
        \label{eq_mu01_defX0}
    \end{equation}
    evaluates $X$ at $x=0$. 
An improper eigenfunction writes
    \begin{eqnarray}
         \dfrac{\breve{v}_\omega (x)}{\vAi}
       = \left\{
         \begin{array}{ll}
            \barke
                 \left[\dfrac{\Ai(X)}{\Ai'(X_0)}-\dfrac{\Bi(X)}{\Bi'(X_0)}
                 \right],   
    		        & \quad 0 \le x \le d, 			\\[0.5cm]
            \Ac \cos(\ke x) + \As \sin(\ke x),   
    				& \quad x > d,
        \end{array} 
     \right.
     \label{eq_mu01_vImproper}
    \end{eqnarray}
    where
    \begin{subequations}
    \label{eq_mu01_ACAS}
    \begin{align} 
    & \Ac =  \barke      Y_1  \cos(\ke d) 
            -\barD^{1/3} Y_1' \sin(\ke d), \\
    & \As =  \barke      Y_1  \sin(\ke d)
            +\barD^{1/3} Y_1' \cos(\ke d).  
    \end{align}
    \end{subequations}
Here $Y_1$ and $Y_1'$ are some coefficients given by
    \begin{subequations}
    \begin{align}
       &Y_0 = \dfrac{\Ai(X_0) }{\Ai'(X_0)}-\dfrac{\Bi(X_0) }{\Bi'(X_0)}, \\
       &Y_1 = \dfrac{\Ai(X_1) }{\Ai'(X_0)}-\dfrac{\Bi(X_1) }{\Bi'(X_0)}, \\
       &Y_1'= \dfrac{\Ai'(X_1)}{\Ai'(X_0)}-\dfrac{\Bi'(X_1)}{\Bi'(X_0)},  
    \end{align}
    \end{subequations}
    with $Y_0$ defined for immediate future use, and
    \begin{equation}
        X_1 = \dfrac{\barkappae^2}{\barD^{2/3}}
        \label{eq_mu01_defX1}
    \end{equation}
    being the value of $X$ evaluated at $x=d$.

Now consider proper eigenmodes. 
An eigenfunction writes
    \begin{eqnarray}
    \dfrac{\breve{v}_j (x)}{\vAi}=  
      \left\{
        \begin{array}{ll}
               \left[\dfrac{\Ai(X)}{\Ai'(X_0)}
                    -\dfrac{\Bi(X)}{\Bi'(X_0)}\right]
               \Exp{-\kappae d},   
    		        & \quad 0 \le x \le d, 			\\[0.5cm]
               \left[\dfrac{\Ai(X_1)}{\Ai'(X_0)}
                    -\dfrac{\Bi(X_1)}{\Bi'(X_0)}\right] 
               \Exp{-\kappae x},   
    				& \quad x > d.
        \end{array} 
     \right.
     \label{eq_mu01_vproper}
    \end{eqnarray}
Some algebra leads to an expression for $\innerp{\hat{v}_j}{\hat{v}_j}$, which reads
    \begin{equation} 
    \label{eq_mu02_innerPproper}
    \begin{split}    
     \dfrac{\innerp{\breve{v}_j}{\breve{v}_j}}{\rhoi \vAi^2 d}
    &  =\dfrac{\Exp{-2\barkappae}}{\barD^{1/3}}
        \left[1-\left(1-\dfrac{\rhoe}{\rhoi}\right)\dfrac{\barki^2}{\barD}
             \right]
        \left[X_1 Y_1^2- (Y_1')^2-X_0 {Y_0}^2
             \right] \\
     &  -\dfrac{\Exp{-2\barkappae}}{\barD^{2/3}}
         \dfrac{1-\rhoe/\rhoi}{3}
         \left[Y_1' Y_1-X_1 (Y_1')^2+X_1^2 Y_1^2-X_0^2 Y_0^2
              \right] \\
    &   + Y_1^2
          \dfrac{\rhoe}{\rhoi}
          \dfrac{\Exp{-2\barkappae}}{2\barkappae}.
      \end{split}  
    \end{equation}
When deriving Eq.~\eqref{eq_mu02_innerPproper}, we have used some identities
    for the indefinite integrals of Airy's functions
    \citep[see Sect.9.11(iv) in][]{NIST:DLMF}.
The DR for proper kink eigenmodes further writes
    \begin{equation}
        \dfrac{\Ai'(X_1) \Bi'(X_0)  -\Ai'(X_0) \Bi'(X_1)}
              {\Ai( X_1)  \Bi'(X_0) -\Ai'(X_0) \Bi( X_1)} 
    = - \dfrac{\barkappae}{\bar{D}^{1/3}}.
        \label{eq_mu01_DR}
    \end{equation}
We note that this DR is not available in the solar literature per se,
    despite the well known applications of Airy's functions in wave contexts
    \citep[e.g.,][]{1983optical..book....S,2018ApJ...855...53L}.   

We proceed to derive the approximate expressions for cutoff wavenumbers $\kcutj$,
    which arise when $\omega/k = \vAe$ 
    and hence $\barkappae=0$ (see Eqs.~\eqref{eq_def_kike} and \eqref{eq_def_barkike}).
It follows from Eq.~\eqref{eq_mu01_defX1} that $X_1=0$, meaning that
    the DR \eqref{eq_mu01_DR} becomes
\begin{equation}
     \dfrac{\Ai'(X_0)}{\Bi'(X_0)} 
   = \dfrac{\Ai'(0)}{\Bi'(0)} 
   = -\dfrac{1}{\sqrt{3}}
   = -\cot\dfrac{\pi}{3}.
   \label{eq_mu01_DRatcutoff}
\end{equation}
Defining 
\begin{equation}
    \zeta \coloneqq \dfrac{2}{3} (-X_0)^{3/2},
\end{equation}
    one finds that $\Ai'(X_0)/\Bi'(X_0)$ is well approximated by 
    $-\cot(\zeta+\pi/4)$ 
    \citep[see Sect.9.7(ii) in][]{NIST:DLMF}.
Note that $\barD=\barki^2=\kcut^2 (\rhoi/\rhoe-1)$ at cutoffs
    (see Eqs.~\eqref{eq_def_kike} and \eqref{eq_def_barkike}).
With $X_0$ now being $-\barD^{1/3}$,
    Eq.~\eqref{eq_mu01_DRatcutoff} then leads to that
    \begin{equation}
               \kcutj d 
       \approx \dfrac{\dfrac{3}{2}\left(j-\dfrac{11}{12}\right)\pi}
                     {\sqrt{\rhoi/\rhoe-1}}, \quad 
            (j=2, 3, \cdots),
       \label{eq_mu01_kcut}  
    \end{equation}
    where the absence of $k_{{\rm cut},1}$ is also made explicit.
This approximate expression is increasingly accurate when $j$ increases,
    overestimating the exact value by merely $0.7\%$ even for $j=2$.
       
\subsection{Further remarks}
\label{sec_sub_solRemarks}

This subsection presents some further remarks that help visualize 
    the mathematical developments so far.
 
\begin{figure}
\centering
\includegraphics[width=.95\columnwidth]{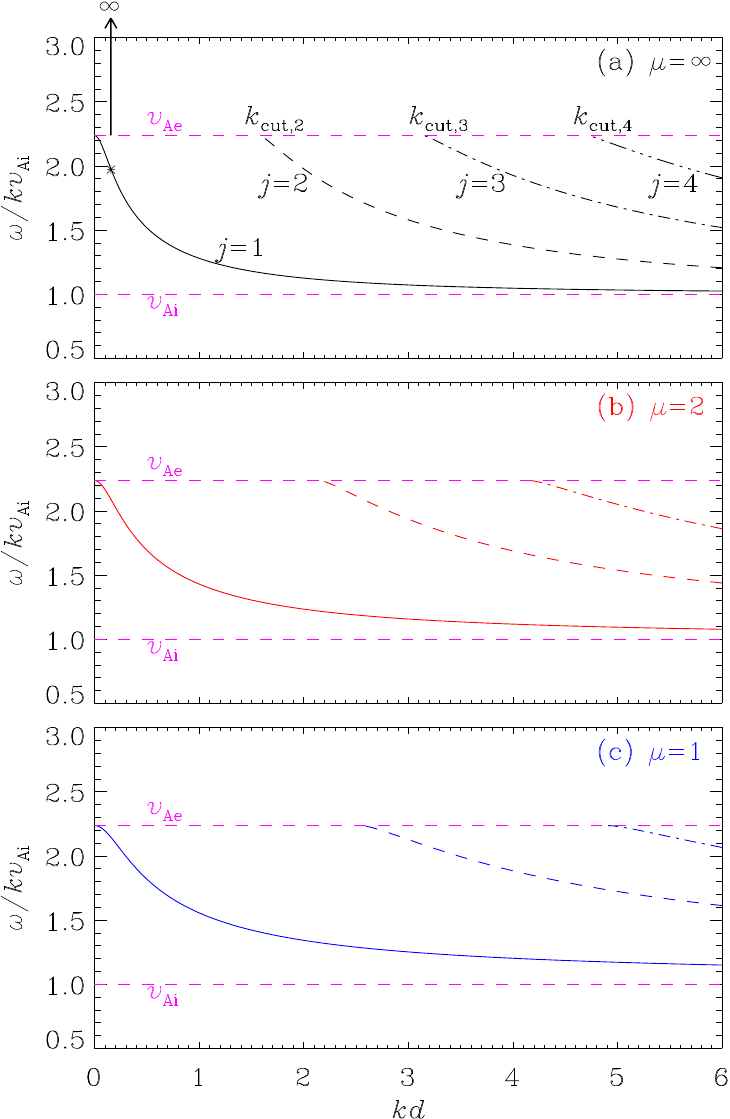}
\caption{
Dependency on the dimensionless axial wavenumber ($kd$) of the dimensionless axial phase
    speed ($\omega/k\vAi$) for the first several branches of proper kink eigenmodes.
The density contrast is fixed at $\rhoi/\rhoe=5$, whereas a number of steepness parameters
    $\mu$ are examined in different panels.
The two horizontal dashed lines represent the internal and external \Alf\ speeds
    ($\vAi$ and $\vAe$). 
Different branches are discriminated by the line styles as illustrated in panel a,
    where our convention for labeling the eigenmodes is also shown.
Cutoff axial wavenumbers arise for all branches except the first one ($j=1$), their labels 
    chosen to be consistent with our mode labeling convention.
Panel a singles out the dimensionless axial wavenumber $kd=\pi/20$ to illustrate 
    the spectrum for the EVP presented in Sect.~\ref{sec_sub_SolOverall}.
The continuum subspectrum is represented by the vertical arrow,  
    the eigenfrequencies $\omega$ continuously extending from $k\vAe$ to infinity.
Only one element is present in the point subspectrum for this chosen $kd$,
    and is represented by the asterisk.
}
\label{fig_DispDiagram_proper} 
\end{figure}

Fixing the density contrast at $\rhoi/\rhoe=5$, 
    Figure~\ref{fig_DispDiagram_proper} presents how the axial phase speed ($\omega/k$) depends on the axial wavenumber ($k$) for the first several branches of proper kink eigenmodes.
Different panels pertain to different values of the steepness parameter $\mu$, 
    with the two horizontal dashed lines representing the internal and external \Alf\ speeds ($\vAi$ and $\vAe$). 
We consistently discriminate different branches by different line styles, 
    and specifically illustrate our mode labeling convention for the step case ($\mu=\infty$, Fig.~\ref{fig_DispDiagram_proper}a).
Also annotated therein are cutoff wavenumbers, which are so labeled as to be
    consistent with the mode labels.   
Note that cutoffs are absent for the first branch.
Note further that the cutoff wavenumber $\kcutj$ tends to increase when $\mu$ decreases
    for a given $\rhoi/\rhoe$ and a given label $j\ge 2$.    
In fact, $k_{{\rm cut}, 4}$ is so large when $\mu=2$ or $\mu=1$ that the $j=4$ branch
    is outside the range for plotting Figs.~\ref{fig_DispDiagram_proper}b
    and \ref{fig_DispDiagram_proper}c.
This behavior of cutoffs $\kcutj$ can be adequately explained by 
    their exact or approximate expressions (e.g., Eq.~\eqref{eq_step_kcut} for $\mu=\infty$).

We proceed to describe the implementation of the EVP-based solution~\eqref{eq_modalSol}. 
To start, we recall that the density contrast $\rhoi/\rhoe$ 
    is allowed to vary between $2$ and $10$, whereas the axial wavenumber
    will be fixed at $kd=\pi/20$.
It follows from Eq.~\eqref{eq_step_kcut} that the inequality
   $kd=\pi/20 < k_{{\rm cut},2}$ consistently holds for the step profile ($\mu=\infty$),
   and therefore holds for other values of $\mu$ as well.
Consequently, only one proper eigenmode is present 
   in the point subspectrum associated with our EVP (see Sect.~\ref{sec_sub_SolOverall}).
Taking the step profile as example, Figure~\ref{fig_DispDiagram_proper}a then displays
   this proper eigenmode as the asterisk. 
Furthermore, the arrow is intended to represent the continuum subspectrum, 
   with the symbol $\infty$ indicating that the eigenfrequency $\omega$ extends
   out to infinity. 
The following steps are then adopted to implement Eq.~\eqref{eq_modalSol}
    for a given combination $[\rhoi/\rhoe, kd, \Lambda/d]$.
\begin{itemize}
\item 
   Given a value of $\mu$, we solve the corresponding DR for the 
       eigenfrequency $\omega_1$ of the only relevant proper eigenmode, 
       and then evaluate the associated eigenfunction $\breve{v}_1$.
   The inner products $\innerp{\breve{v}_1}{\breve{v}_1}$ and $\innerp{u}{\breve{v}_1}$
       are evaluated afterwards, thereby enabling the evaluation of the coefficient $c_1$ (see Eq.~\eqref{eq_Sol_DefcjSomg}).
   The proper contribution is computed with
       the first term on the RHS of Eq.~\eqref{eq_modalSol}, taking $J=1$. 
\item 
   We evaluate the eigenfunctions $\breve{v}_\omega$ for the continuum eigenmodes, 
       obtaining the coefficient $\Ac$ and $\As$ as byproducts. 
   The function $q(\omega)$ is then evaluated with Eq.~\eqref{eq_qomg_generic}, 
       making it straightforward to evaluate the coefficient $S_\omega$ with Eq.~\eqref{eq_Sol_DefcjSomg}).
   The improper contribution is computed with the second
       term on the RHS of Eq.~\eqref{eq_modalSol}.
\end{itemize} 
Despite being essentially analytical, the above steps nonetheless involve some numerical
    evaluations, with the integration over the continuum in Eq.~\eqref{eq_modalSol}
    being an example.
Convergence studies are therefore conducted to ensure that
    no difference can be discerned in our time-dependent solutions when, say, 
    a different grid is employed to discretize the continuum.

\section{EVP-based solutions: specific examples}
\label{sec_results}

\subsection{General spatio-temporal patterns}

\begin{figure}
\centering
\includegraphics[width=.99\columnwidth]{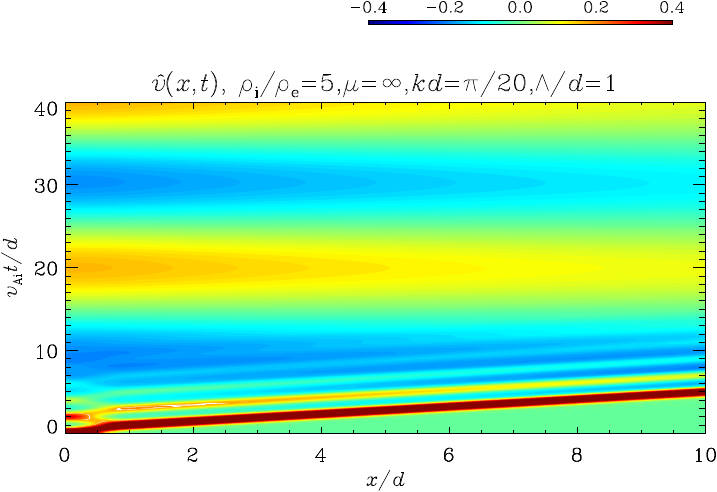}
\caption{
Time-dependent solution obtained with the method of eigenfunction expansion for
    the combination $[\rhoi/\rhoe=5, \mu=\infty, kd=\pi/20, \Lambda/d =1]$.
Shown is the distribution of the lateral speed $\hat{v}$ in the $x-t$ plane, 
    with the filled contours so chosen as to better visualize the wave dynamics
    at short times. 
     }
\label{fig_TDPlot} 
\end{figure}

This subsection presents some generic features of our time-dependent solutions.
We start with a computation pertaining to 
     $[\rhoi/\rhoe=5, \mu=\infty, kd=\pi/20, \Lambda/d =1]$, 
     the lateral speed for which is displayed as a function of $x$ and $t$
     in Fig.~\ref{fig_TDPlot}.
One readily sees that the system evolution comprises two qualitatively different stages.
The short-time response ($t\lesssim 12 d/\vAi$ here)
     to the kink exciter features a series of oblique ridges
     in the slab exterior ($x>d$), signifying partial reflections
     and partial transmissions at the slab-ambient interface. 
By construction, these partial reflections/transmissions are associated
     with the improper contribution in view of the trigonometric
     $x$-dependence of the improper eigenfunctions
     ($\breve{v}_\omega$, Eq.~\eqref{eq_extsol_generic}).
The $x-t$ diagram at large times is then characterized by a series of horizontal stripes, 
     meaning that the external perturbations eventually become laterally standing. 
Evidently, this standing behavior derives from the dominance of the proper contribution;
     see Eq.~\eqref{eq_modalSol} and note that only one proper eigenmode
     is involved in the summation. 
The qualitative behavior in Fig.~\ref{fig_TDPlot} is common to our $t$-dependent solutions.
It is just that partial reflections/transmissions 
     may not be readily recognizable for some combinations of $\rhoi/\rhoe$, $\mu$, and $\Lambda/d$.

\begin{figure}
\centering
\includegraphics[width=.95\columnwidth]{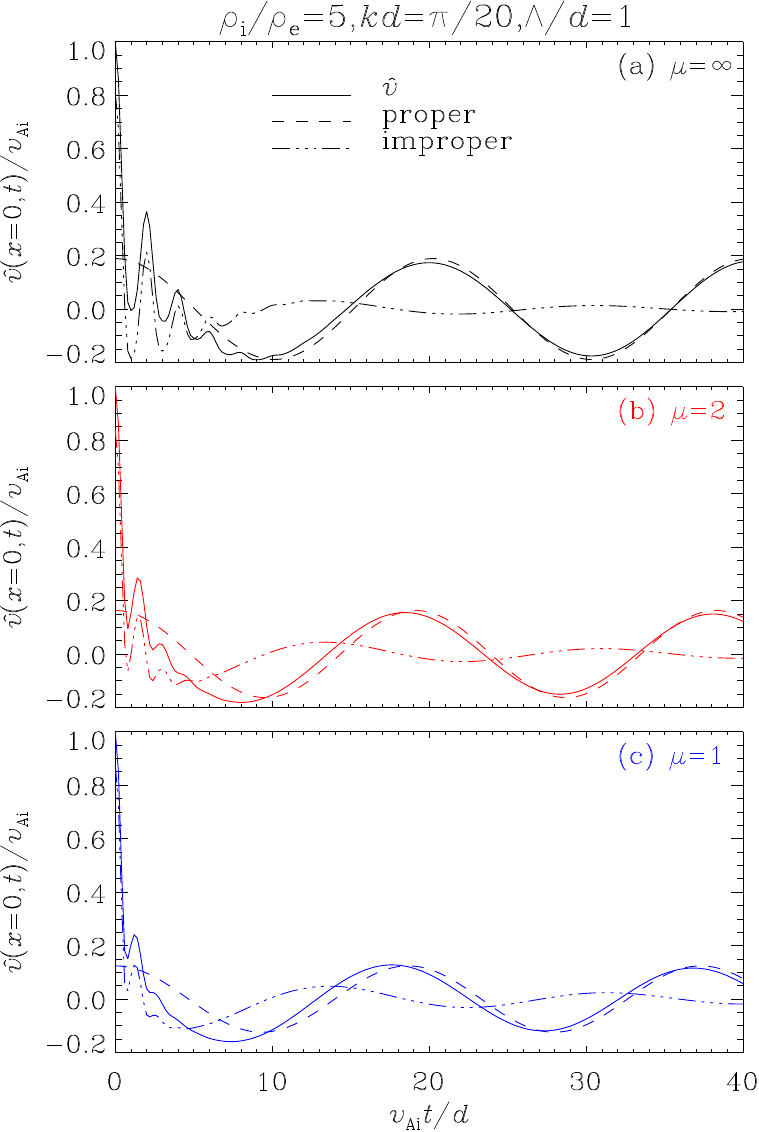}
\caption{
Temporal evolution of the lateral speed at the slab axis $\hat{v}(x=0,t)$. 
A fixed combination $[\rhoi/\rhoe=5, kd=\pi/20, \Lambda/d=1]$ is adopted,
    whereas a number of steepness parameters $\mu$ are examined
    in different panels as labeled. 
The solid curve in each panel shows the lateral speed itself.
The contributions from proper and improper eigenmodes are presented by the dashed
    and dash-triple-dotted curves, respectively.
     }
\label{fig_vAxis_tDep_fixedLambda} 
\end{figure}

We focus on the time sequences of the lateral speed at the slab axis, 
    $\hat{v}(x=0, t)$, from here onward.  
Adopting a fixed combination $[\rhoi/\rhoe=5, kd=\pi/20, \Lambda/d=1]$,
    Figure~\ref{fig_vAxis_tDep_fixedLambda} displays the $\hat{v}(x=0, t)$ profiles
    for different steepness parameters $\mu$ in different panels as labeled.  
In addition to $\hat{v}(x=0, t)$ itself (the solid curve in each panel), 
    the contributions from proper and improper eigenmodes are further displayed  
    by the dashed and dash-triple-dotted curves, respectively.
Once again, one readily discerns the two-stage behavior.
Let us first consider the proper contribution, namely 
    a simple sinusoid with period $P_1 = 2\pi/\omega_1$
    (see Eq.~\eqref{eq_modalSol}).  
We recall that $\omega_1$ is only slightly smaller than $\omgcrit=k\vAe$ 
    (see Fig.~\ref{fig_DispDiagram_proper}), meaning that $P_1$ is only slightly longer
    than the axial \Alf\ time $2\pi/\omgcrit=2L/\vAe$ ($\approx 17.9d/\vAi$ here) 
    for all the examined steepness parameters.
We recall further that the proper contribution is stationary
    in magnitude (see Eq.~\eqref{eq_modalSol}), meaning that any overall two-stage behavior in $\hat{v}(x=0, t)$ is carried by the improper contribution.
For any dash-triple-dotted curve, one then sees a transition from some rapid attenuation
    at short times to some later stage where the temporal attenuation
    is substantially slower.  
A comparison between different panels further indicates that 
    this transition tends to occur later for larger $\mu$.
One also sees that some short periodicities on the order of 
    the lateral \Alf\ time ($d/\vAi$) 
    may be present in the initial stage (e.g., Fig.~\ref{fig_vAxis_tDep_fixedLambda}a), and these short periodicities tend to persist longer for large $\mu$ (compare e.g., Fig.~\ref{fig_vAxis_tDep_fixedLambda}a with Fig.~\ref{fig_vAxis_tDep_fixedLambda}c).
Regardless, the improper contribution at large times consistently
    features a longer periodicity that is only marginally below   
    $2\pi/\omgcrit=2L/\vAe$.

The temporal behavior in the improper contribution can be understood 
    from two complementary perspectives.
The first perspective, based on partial reflection/transmission, is visually more intuitive for 
    interpreting the short periodicities ($\sim d/\vAi$) at early times (see also Fig.~\ref{fig_TDPlot}).
Energetically, this perspective is also more intuitive
    in understanding the overall tendency for the improper contribution to diminish with time because this tendency can take place only via the incessant transmission of fast perturbations into the ambient fluid.
Our second perspective, built directly on Eq.~\eqref{eq_modalSol}, is to attribute all the temporal features 
    to the interference among the continuum eigenmodes. 
For instance, the temporal attenuation is connected to the effect whereby 
    any adjacent monochromatic components of the continuum will become increasingly out-of-phase as time proceeds.
Physically, this effect is nothing but destructive interference;
    the increasingly out-of-phase cosines in Eq.~\eqref{eq_modalSol} 
    tend to cancel out. 
One then expects that the low-frequency portion with 
    $\omega \gtrsim \omgcrit=k\vAe$ will gradually stand out, because it takes longer for the improper eigenmodes in this portion to get out-of-phase.
This expectation is reproduced by Fig.~\ref{fig_vAxis_tDep_fixedLambda} 
    where any dash-triple-dotted curve
    is eventually characterized by some slow attenuation 
    and some periodicity  $\lesssim 2\pi/\omgcrit$.
One also expects that the contribution from the high-frequency portion will
    attenuate more rapidly, 
    which is indeed seen at early times.
That this initial stage involves high-frequency improper eigenmodes is also corroborated by 
    the short periodicities $\sim d/\vAi$ therein.  

\begin{figure*}
\centering
\includegraphics[width=.95\textwidth]{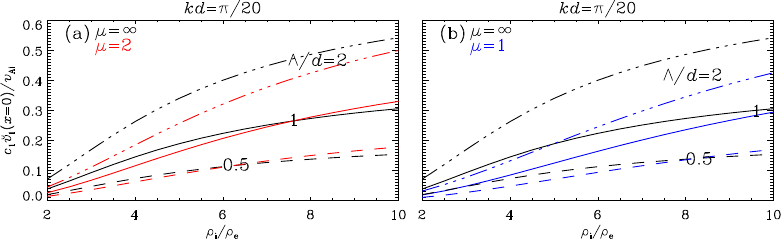}
\caption{
Dependence of $c_1 \breve{v}_1(x=0)$ on the density contrast $\rhoi/\rhoe$ 
    for several values of $\Lambda/d$ as labeled and discriminated by linestyles. 
The results for the step case ($\mu=\infty$, the black curves)
    are compared with those for (a) $\mu=2$ (red) and (b) $\mu=1$ (blue).
The axial wavenumber is fixed at $kd=\pi/20$.
Note that $c_1 \breve{v}_1(x=0)$ measures the magnitude of the asymptotic variation
    at the slab axis; see text for details.
}
\label{fig_vAxis_c1v1} 
\end{figure*}

\begin{figure*}
\centering
\includegraphics[width=.95\textwidth]{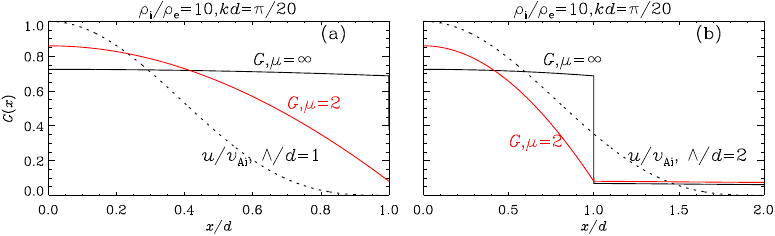}
\caption{
The function $G(x)$ plotted as solid curves over the ranges (a) $[0, d]$ and (b) $[0, 2d]$
    for two different steepness parameters $\mu$ as discriminated by different colors.
Plotted by the dotted curves are two initial perturbations $u(x)$
    with (a) $\Lambda/d=1$ and (b) $\Lambda/d=2$.
A fixed combination $[\rhoi/\rhoe=10, kd=\pi/20]$ is employed.
See text for details.}
\label{fig_vAxis_Gx} 
\end{figure*}

The interference perspective is applicable to broader solar contexts.
We consider only those studies where the eigenfunction expansion method is adopted.
We further restrict ourselves to those time intervals 
    where some perturbation at a given position 
    decays with time as a result of the destructive interference 
    of continuum eigenmodes. 
Evidently, this interference perspective holds regardless
    of the origin of the continuum. 
We then discriminate two situations, in one of which the continuum results
    from the system being laterally open and in the other the continuum arises because
    of some continuous transverse structuring inside the domain. 
Our study is an example for the former.
The physics is identical to what happens to wave motions
    in equilibrium setups with step transverse
    profiles for cylindrical \citep{2014ApJ...789...48O,2015ApJ...806...56O,2022ApJ...928...33L} and slab geometries alike (\citealt{2023ApJ...943...91W}; see also \citealt{2007PhPl...14e2101A}).
What our study demonstrates is that a further physically relevant continuum
    does not necessarily arise when
    the transverse structuring is continuous. 
When indeed physically relevant, however, the pertinent continuum may play a vital role in explaining, say, 
     the intricate dependence on the equilibrium quantities of damping envelopes of kink motions in coronal cylinders (\citealt{2015ApJ...803...43S}; see also \citealt{1991JPlPh..45..453C}). 
On this aspect we note that the study by \citet{2015ApJ...803...43S} focused on the resonant damping
     in the \Alf\ continuum, and hence discarded the continuum
     that is introduced by the system being unbounded.  
To our knowledge, the eigenfunction expansion method remains to be applied to the situation
     that physically involves continua of distinct origins.

\subsection{Contribution from proper eigenmodes}

This subsection focuses on the proper contribution, the purpose being
    to illustrate some subtleties 
    in assessing the capabilities for the examined equilibria to confine the energy imparted by the initial exciter. 
An intuitive indicator will be the dimensionless axial cutoff wavenumber $\kcutj d$
    for any given label $j$, 
    and we take $k_{{\rm cut},2} d$ for the ease of description.
The intuition behind is that a smaller $k_{{\rm cut},2} d$ 
    represents some stronger confinement capability.
Note that $\kcutj d \sqrt{\rhoi/\rhoe-1}$ for a given $j$ depends only
    on the function $f(x)$
    for any equilibrium density distribution describable by Eq.~\eqref{eq_prof_gen}, as explained by \citet{2018ApJ...855...53L} even though sausage proper eigenmodes 
    were examined therein.         
Specializing to our inner-$\mu$ profile (Eq.~\eqref{eq_prof_innermu}), one recognizes that
    $k_{{\rm cut},2} d \sqrt{\rhoi/\rhoe-1}$ depends only on the steepness parameter $\mu$.
It then follows that $k_{{\rm cut},2} d$ decreases with $\rhoi/\rhoe$ or $\mu$ at least for
    the examined steepness parameters (see Eqs.~\eqref{eq_step_kcut}, \eqref{eq_mu02_kcut}, and \eqref{eq_mu01_kcut}). 
This is indeed intuitive given that a slab deviates more strongly from its ambient fluid
    when $\rhoi/\rhoe$ or $\mu$ increases, and hence one expects a stronger proper contribution inside the slab.
We focus on the slab axis, and take $c_1 \breve{v}_1(x=0)$ 
    to illustrate that this expectation
    does not necessarily hold.

Figure~\ref{fig_vAxis_c1v1} displays $c_1 \breve{v}_1(x=0)$ as a function of 
    the density contrast $\rhoi/\rhoe$ for several $\Lambda/d$ 
    as discriminated by the different linestyles.
The specific values of $\Lambda/d$ are further labeled. 
In addition, the results for the step case ($\mu=\infty$, the black curves)
    are compared with those for $\mu=2$ (red, see Fig.~\ref{fig_vAxis_c1v1}a)
    and $\mu=1$ (blue, Fig.~\ref{fig_vAxis_c1v1}b). 
Examining any curve, one sees that $c_1 \breve{v}_1(x=0)$ consistently increases with 
    $\rhoi/\rhoe$, as expected for some stronger structuring. 
The $\mu$-dependence, on the other hand, is more subtle if one inspects 
    a pair of curves with the same linestyle but different colors. 
The calculation for a larger $\mu$ may indeed yield a larger $c_1 \breve{v}_1(x=0)$
    when the initial exciter is more spatially extended
    (e.g., the dash-triple-dotted curves in Fig.~\ref{fig_vAxis_c1v1}a labeled $\Lambda/d=2$).
However, the opposite may also take place as indicated by, say,  
    the rightmost portions of the curves labeled $\Lambda/d=1$ in Fig.~\ref{fig_vAxis_c1v1}a.

Why does the somehow counterintuitive $\mu$-dependence occur occasionally?
We address this by capitalizing on Eqs.~\eqref{eq_Sol_DefcjSomg} and \eqref{eq_def_innerprod}
    to rewrite $c_1 \breve{v}_1(x=0)$ as
    \begin{equation}
        \label{eq_figRef_c1v1}
          c_1 \breve{v}_1(x=0) 
        = \int\limits_{0}^{\Lambda} u(x) \dfrac{G(x)}{d} \mathd x,
    \end{equation}    
    where
    \begin{equation}
      \label{eq_figRef_Gx}
       G(x) 
       \coloneqq  
                        \dfrac{[\rho_0(x) \breve{v}_1(x)]  \breve{v}_1(x=0)}
             {\int_{0}^{\infty} \rho_0(x) \breve{v}^2_1(x) \mathd x}
             d. 
    \end{equation}
The introduction of the slab half-width $d$ into
    Eqs.~\eqref{eq_figRef_c1v1} and \eqref{eq_figRef_Gx} is immaterial, the purpose
    being simply to make $G(x)$ dimensionless. 
Defined this way, $G(x)$ allows the proper eigenfunction to be arbitrarily scaled.
One may therefore interpret $G(x)$ in Eq.\eqref{eq_figRef_c1v1} as some response function
    of a system that is fully determined by $[\rhoi/\rhoe, \mu, kd]$.
The specific output $c_1 \breve{v}_1(x=0)$ is then determined by how 
    the input $u(x)$ is distributed over the $G(x)$ profile. 
We arbitrarily specialize to a combination $[\rhoi/\rhoe=10, kd=\pi/20]$. 
The response function $G(x)$ then depends only on $\mu$, and the associated profiles
    are plotted in Fig.~\ref{fig_vAxis_Gx} for two different values of $\mu$,
    one being $\mu=\infty$ (the black solid curves) and the other being $\mu=2$ (red).
Note that Figs.~\ref{fig_vAxis_Gx}a and \ref{fig_vAxis_Gx}b differ only in 
    the range of the horizontal axis as far as the solid curves are concerned.     
The $u(x)$ profile is additionally plotted by the dotted curve for (a) $\Lambda/d=1$
    and (b) $\Lambda/d=2$, and we note that $u(x)$ is independent of $G(x)$.
A comparison between the two $G(x)$ profiles 
    shows that they differ primarily in that $G(x)$ for $\mu=2$ 
    exceeds its $\mu=\infty$ counterpart in the interval $x\lesssim 0.4 d$. 
The integral in Eq.~\eqref{eq_figRef_c1v1} then means that this portion weighs more when 
    the initial exciter $u(x)$ is more spatially localized, making $c_v \breve{v}_1(x=0)$
    larger for $\mu=2$ rather than for $\mu=\infty$. 
This pertains to Fig.~\ref{fig_vAxis_Gx}a where $\Lambda/d=1$.
However, the portion $x\gtrsim 0.4 d$ plays a more important role
    when $u(x)$ is sufficiently extended, 
    thereby explaining why $c_v \breve{v}_1(x=0)$ for $\mu=\infty$
    is larger as happens when $\Lambda/d=2$ (see Fig.~\ref{fig_vAxis_Gx}b).
One may question that the subtle $\mu$-dependence
    in Fig.~\ref{fig_vAxis_c1v1} holds only when $u(x)$ takes the specific form of Eq.~\eqref{eq_u}, which may indeed be true. 
However, our point is that the interpretation of $G(x)$ as a response function
    provides a sufficiently general framework for understanding the system responses to different implementations of $u(x)$.
Furthermore, with the subtle $\mu$-dependence we actually highlight the 
    importance of the details of the initial exciter $u(x)$
    for determining how the system responds.   

\subsection{Contribution from improper eigenmodes}

This subsection examines the improper contribution, 
     again specializing to the slab axis ($x=0$).  
Let $\hat{v}_{\rm improper}(x,t)$ denote the integral in Eq.~\eqref{eq_modalSol}.
This subsection focuses specifically on the connection between the short periodicities 
     at early times in $\hat{v}_{\rm improper}(x=0,t)$ and the pertinent DLM expectations. 
An examination on DLMs is therefore necessary in the framework of classic BVPs, for which
     we recall that the concept of DRs always applies.
We present such an examination in Appendix~\ref{sec_App_DLM},
     where the DRs for continuous profiles are new in solar contexts.
A countable infinity of modes arise
     for any axial wavenumber $kd$ when given a combination $[\rhoi/\rhoe, \mu]$. 
Let $\Omega_j$ be the frequency of the $j$-th mode, and see its real part $\Re\Omega_j$
     as positive without loss of generality. 
By DLMs we refer to the infinite subset of modes
     with the defining features that $\Re\Omega_j > k\vAe$ and $\Im\Omega_j < 0$.
In our study, all modes with $j\ge 2$ quality as DLMs 
     for all $[\rhoi/\rhoe, \mu]$ given the chosen $kd=\pi/20$. 
We nonetheless denote any such frequency as $\Omega^{\rm DLM}_{j}$ 
     to emphasize the distinction between the oscillation frequency 
     $\Re\Omega^{\rm DLM}_{j}$ and any (real-valued) eigenfrequency $\omega$ in our improper continuum.

\begin{figure}
\centering
\includegraphics[width=.99\columnwidth]{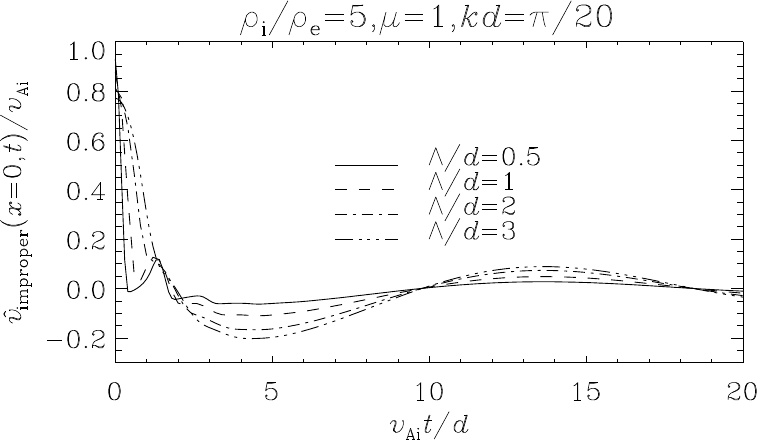}
\caption{
Temporal evolution of the improper contribution as given by the integral in Eq.~\eqref{eq_modalSol}. 
A fixed combination $[\rhoi/\rhoe=5, \mu=1, kd=\pi/20]$ is adopted, 
    whereas a number of values are examined for $\Lambda/d$ as labeled.}
\label{fig_vAxis_vImproper} 
\end{figure}

\begin{figure}
\centering
\includegraphics[width=.98\columnwidth]{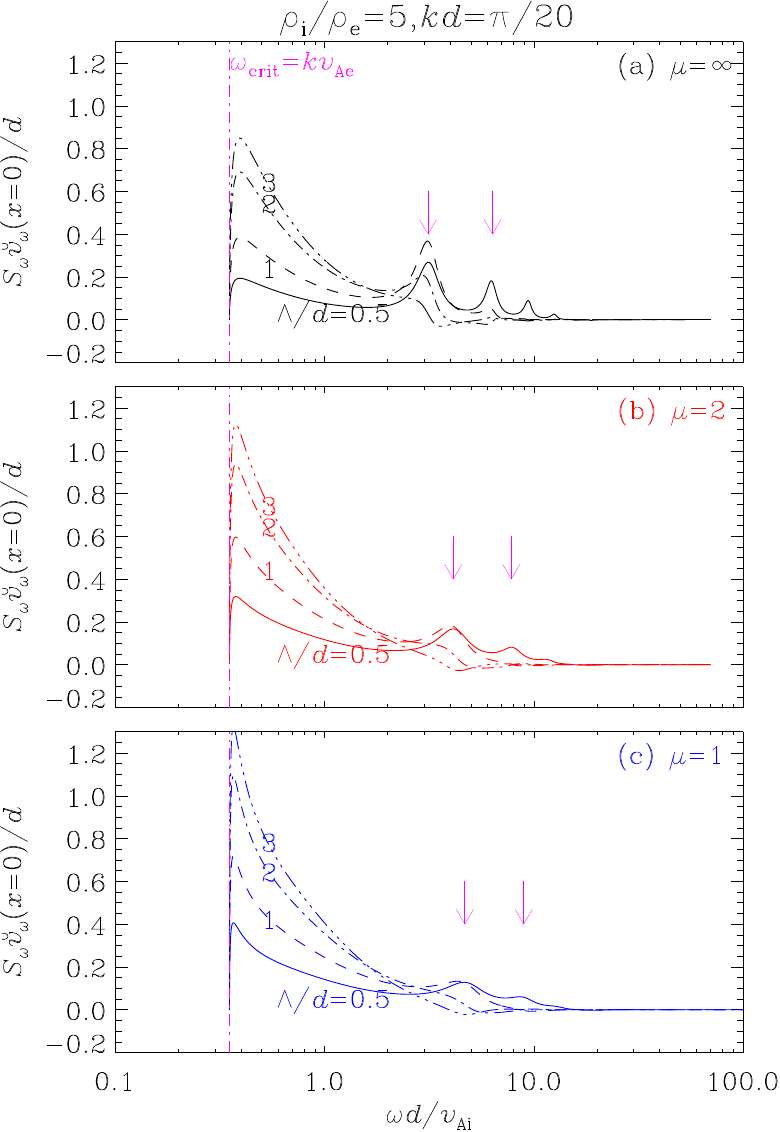}
\caption{
Frequency-dependencies of the local spectral density evaluated at the slab axis,
  $S_\omega \breve{v}_\omega(x=0)$.
A fixed combination $[\rhoi/\rhoe=5, kd=\pi/20]$ is adopted, whereas 
    a number of values are examined for the steepness parameter $\mu$ 
    in different panels. 
For each $\mu$, a number of values of $\Lambda/d$ are experimented with
    and discriminated by the linestyles. 
The magenta arrows in each panel mark the oscillation frequencies 
    of the first two discrete leaky modes
    ($\Re\Omega^{\rm DLM}_{2}$ and $\Re\Omega^{\rm DLM}_{3}$)
    computed for the same set of $[\rhoi/\rhoe, \mu, kd]$.
All vertical dash-dotted lines correspond to the critical frequency
    $\omgcrit=k\vAe$.
See text for more details.    
    }
\label{fig_vAxis_SomgVomg} 
\end{figure}

Some general expectations follow from Eq.~\eqref{eq_modalSol}.
We start by reformulating $S_\omega \breve{v}_\omega(x=0)$ as 
    \begin{equation}
        \label{eq_figRef_SomgVomg}    
        S_\omega \breve{v}_\omega(x=0)
      = \int\limits_0^{\Lambda} u(x) G_\omega(x) \mathd x,
    \end{equation}
    where 
    \begin{equation}
     \label{eq_figRef_Gomg}
           G_\omega(x) 
   \coloneqq  \dfrac{\left[\rho_0(x) \breve{v}_\omega (x)\right] \breve{v}_\omega (x=0)}
                    {q(\omega)},
    \end{equation}
    and we have used Eqs.~\eqref{eq_def_innerprod} and \eqref{eq_Sol_DefcjSomg}.
Equation~\eqref{eq_figRef_SomgVomg} closely resembles Eq.~\eqref{eq_figRef_c1v1}, meaning
    that $G_\omega(x)$ can also be interpreted as some response function.
Note that neither $G_\omega(x)$ nor $\Re\Omega^{\rm DLM}_{j}$ depends on $u(x)$, 
    and suppose for now that the combination $[\rhoi/\rhoe, \mu, kd]$ is given.  
One then expects that some elements in the set $\{\Re\Omega^{\rm DLM}_{j}\}$
    are bound to appear for some $u(x)$ provided that this set is special 
    regarding the $\omega$-dependence of $G_\omega(x)$. 
It is just that whether this expectation holds depends on the details of $u(x)$, which 
    is solely characterized by the spatial extent $\Lambda$ here.

Figure~\ref{fig_vAxis_vImproper} takes a fixed $[\rhoi/\rhoe=5, \mu=1, kd=\pi/20]$
    to illustrate what happens to $\hat{v}_{\rm improper}(x=0, t)$ when $\Lambda/d$ varies as labeled. 
Two features are evident.  
Firstly, some long periodicity $\sim 2\pi/k\vAe$ can always be identified at large times,
    and tends to be more prominent for larger $\Lambda$.
Secondly, and more importantly, short periodicities $\sim d/\vAi$ do show up at early times
    provided that $\Lambda/d$ is sufficiently small, one example being the solid line that pertains to $\Lambda/d=0.5$. 
Figure~\ref{fig_vAxis_SomgVomg}c then address why these features arise 
    by displaying the local spectral density $S_\omega \breve{v}_\omega(x=0)$
    as a function of $\omega$, discriminating different $\Lambda/d$ by the different linestyles. 
The vertical dash-dotted line in magenta represents the critical frequency
    $\omgcrit=k\vAe$.
Note that $S_\omega \breve{v}_\omega(x=0)$ approaches zero 
    when $\omega$ approaches $\omgcrit$ from above or equivalently when $\barke \to 0^+$ (see Eq.~\eqref{eq_def_kike}).
As can be readily verified, the denominator $q(\omega)$
    on the RHS of Eq.~\eqref{eq_figRef_Gomg} scales as $\barke$ whereas 
    the numerator scales as $\barke^2$ in this situation. 
Figure~\ref{fig_vAxis_SomgVomg}c then indicates that a peak always stands out
    in $S_\omega \breve{v}_\omega(x=0)$ (or equivalently its modulus) at some low frequency
    close to $\omgcrit$ and this peak strengthens with $\Lambda/d$, 
    thereby explaining both the persistence of the long periodicity and its $\Lambda$-dependence. 
On the other hand, more peaks at frequencies substantially higher than $\omgcrit$
    become increasingly prominent when $\Lambda/d$ decreases.
This naturally accounts for the behavior of the short periodicities
    in Fig.~\ref{fig_vAxis_vImproper}.
On top of that, the high-frequency peaks become increasing
    close to the oscillation frequencies expected for the DLMs, among which 
    the first two ($\Re\Omega^{\rm DLM}_{2}$ and $\Re\Omega^{\rm DLM}_{3}$) are marked by the magenta arrows.
Now move on to the results for (a) $\mu=\infty$ and (b) $\mu=2$.
All qualitative features in Fig.~\ref{fig_vAxis_SomgVomg}c 
    are seen to hold for these cases as well. 
Some quantitative differences arise nonetheless, one example being that
    the short periodicities are easier to develop for a larger $\mu$
    (compare the solid curve in Fig.~\ref{fig_vAxis_SomgVomg}c with that in, say, Fig.~\ref{fig_vAxis_SomgVomg}a).

\begin{figure}
\centering
\includegraphics[width=.98\columnwidth]{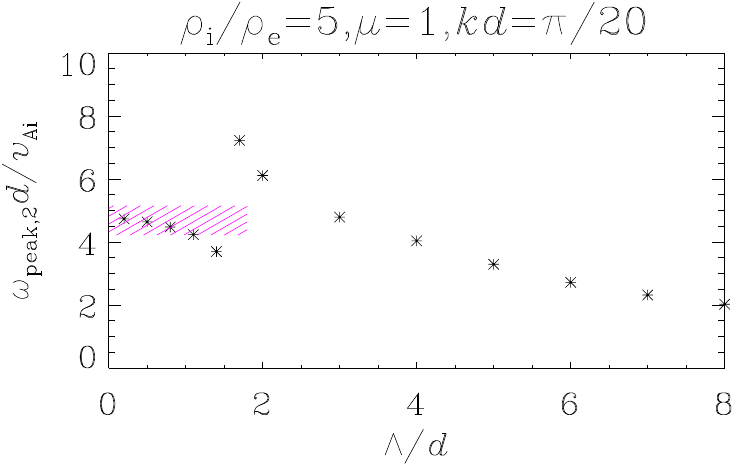}
\caption{
Dependence on $\Lambda/d$ of the frequency $\omega_{{\rm peak}, 2}$
    where $|S_\omega \breve{v}_\omega(x=0)|$ attains the second peak. 
A fixed combination $[\rhoi/\rhoe=5, \mu=1, kd=\pi/20]$ is adopted. 
The hatched portion corresponds to the frequency 
    range $[0.9, 1.1] \times \Re\Omega^{\rm DLM}_{2}$. 
See text for details. 
}
\label{fig_vAxis_trackpeak2} 
\end{figure}

What makes $\Re\Omega^{\rm DLM}_{j}$ special in the improper contribution?
In view of Eq.~\eqref{eq_figRef_SomgVomg}, 
   the most likely reason for the systematic $\Lambda$-dependence of
   the high-frequency peaks is that the set $\{\Re\Omega^{\rm DLM}_{j}, j\ge 2\}$ is special for the response function $G_\omega(x)$.
However, why this happens remains puzzling because the DLMs pertain to
   a classic BVP, which observes a different BC from our IVP.
It turns out that the function $q(\omega)$ plays a decisive role in determining 
   the $\omega$-dependence of $G_\omega(x)$ (see Eq.~\eqref{eq_figRef_Gomg}). 
We illustrate this point by considering the simplest case
   $\mu=\infty$, noting that similar arguments can be offered for other values of $\mu$.
To start, we note that the inequality $|\Omega^{\rm DLM}_{j}|^2 \gg k^2 \vAe^2$ 
   consistently holds in our study. 
When $\mu=\infty$, the DR pertinent to DLMs (Eq.~\eqref{eq_step_classicDR}) 
   then dictates that 
   \begin{equation}
       \label{eq_figRef_ReOmgDLM}
       \Re\Omega^{\rm DLM}_{j}d/\vAi \approx (j-1)\pi,
       \quad j=2, 3, \cdots,
   \end{equation}
   as was first shown for the limiting case $kd \to 0$ by \citet{2005A&A...441..371T}.
Now consider the portion of the improper continuum
   that satisfies $\omega^2 \gg k^2 \vAe^2 > k^2 \vAi^2$ and hence 
   $\ki^2 \approx \omega^2/\vAi^2, \ke^2 \approx \omega^2/\vAe^2$
   (see Eq.~\eqref{eq_def_kike}).
The expressions for the coefficients $\Ac$ and $\As$ 
   (see Eq.~\eqref{eq_step_ACAS}) lead to that
   \begin{equation}
       \label{eq_figRef_sumAc2As2}
            \Ac^2+\As^2 
    \approx \dfrac{\rhoe}{\rhoi} 
          + \left(1-\dfrac{\rhoe}{\rhoi}\right) 
               \sin^2\left(\dfrac{\omega d}{\vAi}\right). 
   \end{equation}
Evidently, the values of $\Re\Omega^{\rm DLM}_{j}$ minimize $\Ac^2+\As^2$
   and hence tend to minimize $q(\omega)$ (see Eq.~\eqref{eq_qomg_generic}), meaning
   that these values tend to stand out as extrema in the response function $G_\omega(x)$
   and hence in the output $S_\omega\breve{v}_\omega(x=0)$
   (see Eqs.~\eqref{eq_figRef_SomgVomg} and \eqref{eq_figRef_Gomg}). 
Equation~\eqref{eq_figRef_sumAc2As2} is reminiscent of Eq.(83) 
   in \citet{2007PhPl...14e2101A} who addressed a similar IVP with the more involved
   Laplace transform approach.
Equation~\eqref{eq_figRef_sumAc2As2} bears some even closer resemblance to
   Eq.~(39) in \citet{2023ApJ...943...91W} where we examined the temporal responses
   to sausage exciters with the same eigenfunction expansion method.
 
At this point, we reiterate that IVP studies in general offer a fuller picture
   for the wave dynamics than classic mode analyses.
As concrete examples, our Figs.~\ref{fig_vAxis_vImproper} and \ref{fig_vAxis_SomgVomg} 
   illustrate that the DLM expectations do not necessarily materialize. 
That said, the concept of DLMs may still prove useful as a shortcut
   approach for interpreting the complicated system evolution in appropriate
   spatio-temporal domains. 
By ``shortcut'' we refer to the general fact that classic mode analyses are 
   computationally much cheaper than IVP studies.
Furthermore, the theoretical results for DLMs are independent from initial exciters
   and hence are much easier to implement than IVP studies from the seismological standpoint.
It therefore should be informative to explore the requirements on the initial exciter
   ($\Lambda/d$ here) for the concept of DLMs to make practical sense. 

Figure~\ref{fig_vAxis_trackpeak2} takes $[\rhoi/\rhoe=5, \mu=1, kd=\pi/20]$ 
    to illustrate how we quantify the range of $\Lambda/d$ 
    where the concept of DLMs is useful.
Basically, we focus on the $\omega$-dependence of $|S_\omega\breve{v}_\omega(x=0)|$, and
    systematically reduce $\Lambda/d$ to track the variation of the position of the second peak ($\omega_{{\rm peak}, 2}$).
Note that $|S_\omega\breve{v}_\omega(x=0)|$ always features a peak
    at some frequency $\gtrsim \omgcrit=k\vAe$, and we see this peak as the first one. 
Figure~\ref{fig_vAxis_trackpeak2} shows a discontinuity around
    some sufficiently small $\Lambda/d$, which is to be denoted $(\Lambda/d)_1$
    and reads $\sim 1.5$ in this particular case.
This behavior is actually common to all of our computations, and results from our 
    convention for numbering the peaks. 
Specifically, it always holds that the second peak initially 
    moves toward higher frequencies when 
    $\Lambda/d$ decreases toward $(\Lambda/d)_1$.
When $\Lambda/d$ decreases further, however, an additional peak becomes identifiable
   at some lower frequency and therefore qualifies as our second peak. 
The value $\omega_{{\rm peak}, 2}$ then increases with decreasing $\Lambda/d$ and 
    eventually approaches $\Re\Omega^{\rm DLM}_{2}$.
Let $(\Lambda/d)_{\rm crit}$ denote the value of $\Lambda/d$ where
    $\omega_{{\rm peak}, 2}$ equals $0.9 \Re\Omega^{\rm DLM}_{2}$.
Equivalently, $(\Lambda/d)_{\rm crit}$ is where
    the $\Lambda-\omega_{{\rm peak}, 2}$ curve in 
    Fig.~\ref{fig_vAxis_trackpeak2} enters the hatched area, for which the lower edge 
    is set to be $0.9 \Re\Omega^{\rm DLM}_{2}$ and the upper edge is arbitrarily
    taken to be $1.1 \Re\Omega^{\rm DLM}_{2}$.
We deem the portion $\Lambda/d \le (\Lambda/d)_{\rm crit}$ as where
    the concept of DLMs helps.  
Here the specific threshold factor $0.9$ is not that important; taking 
    $(\Lambda/d)_{1}$ as $(\Lambda/d)_{\rm crit}$ serves our purposes equally well.
One may question that the frequency range delineated by the hatched area
    is also crossed by the $\Lambda-\omega_{{\rm peak}, 2}$ curve in the portion
    $\Lambda/d > (\Lambda/d)_{1}$. 
The reason for us to discard this large-$\Lambda$ portion is that
    additional peaks emerge when $\Lambda/d$ decreases from $(\Lambda/d)_1$.
Once identifiable, the positions of these peaks become almost instantly close to 
    the expectations for additional DLMs ($\Re\Omega^{\rm DLM}_{j}$ with $j\ge 3$).
This can be readily seen by comparing, say, the solid and dashed curves
    in Fig.~\ref{fig_vAxis_SomgVomg}.
In other words, the occasional proximity of $\omega_{{\rm peak}, 2}$ 
    to $\Re\Omega^{\rm DLM}_{2}$ for some $\Lambda/d > (\Lambda/d)_1$ 
    is actually irrelevant. 

\begin{figure}
\centering
\includegraphics[width=.95\columnwidth]{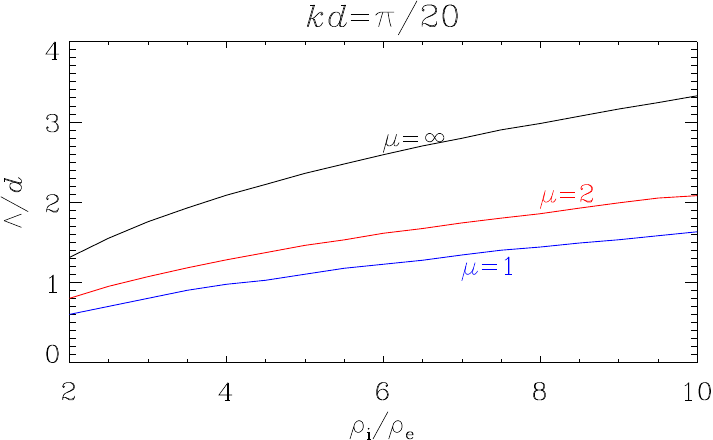}
\caption{
Dividing curves in the $\rhoi/\rhoe-\Lambda/d$ plane separating where
     the concept of discrete leaky modes helps (the portion below a curve)
     from where it does not (above).
The axial wavenumber is fixed at $kd=\pi/20$, whereas a number of 
     steepness parameters $\mu$ are examined as discriminated by the different colors. 
}
\label{fig_vAxis_critLambda} 
\end{figure}

Figure~\ref{fig_vAxis_critLambda} displays $(\Lambda/d)_{\rm crit}$ 
    as a function of the density contrast $\rhoi/\rhoe$ for a number of steepness parameters $\mu$ as labeled. 
The axial wavenumber is fixed at $kd=\pi/20$. 
Evidently, the curve for each $\mu$ serves as a dividing line in the     
    $\rhoi/\rhoe-\Lambda/d$ plane, with the concept of DLMs being helpful only 
    in the portion below the curve. 
Now see $(\Lambda/d)_{\rm crit}$ as a function of $\rhoi/\rhoe$ and $\mu$.
Figure~\ref{fig_vAxis_critLambda} then indicates that $(\Lambda/d)_{\rm crit}$
    increases monotonically with $\rhoi/\rhoe$ or $\mu$ when the other parameter is fixed.
Put together, this agress with the intuition that 
    the concept of DLMs
    may help interpret the system evolution for a broader range of initial exciters
    provided that a slab makes a stronger distinction from its surroundings.
However, $(\Lambda/d)_{\rm crit}$ is consistently smaller than $\sim 3.3$
    for density contrasts representative of AR loops.
The DLM theories therefore seem to be practically useful
    only for those situations where the exciters laterally span 
    no more than several $10^{3}~{\rm km}$, given typical widths of AR loops. 
We argue that this spatial extent is not that small when placed in 
    the context of, say, energy-release processes in solar flares
    \citep[see e.g., the review by][]{2011LRSP....8....6S}.
In our opinion, it is more of an issue that the sort periodicities need
    to be resolved with a cadence higher than typically 
    implemented for, say, EUV imagers.

\section{Summary}
\label{sec_conc} 
This study was largely intended to address, in a physically transparent manner,
     how classic mode analyses 
     connect to the time-dependent wave behavior in structured media.
We chose to work in linear ideal MHD for simplicity, and examined
     axial fundamental kink motions that ensue when localized velocity perturbations are introduced to some symmetric, straight, field-aligned slab equilibria.
Only two-dimensional (2D) motions were of interest, and the equiliria were taken to be 
     structured only transversely.
Continuous structuring was allowed for inside the slab, while the slab exterior was assumed
     to be uniform (Eqs.~\eqref{eq_prof_gen} and \eqref{eq_prof_innermu}).
A 1D initial value problem (IVP) was formulated for laterally open systems, with no 
     definitive requirement imposed at infinity.
An eigenvalue problem (EVP) was constructed correspondingly, enabling the IVP to be 
     analytically solved in terms of the complete set of eigenfunctions. 
This $t$-dependent solution, given by Eq.~\eqref{eq_modalSol}, allows a clear distinction
     between the contribution from proper eigenmodes and that due to improper eigenmodes.
The wave dynamics depends on two subgroups of parameters. 
One subgroup characterizes the initial exciter 
    (the dimensionless spatial extent $\Lambda/d$ here, Eq.~\eqref{eq_u}).
The other subgroup characterizes the equilibria, involving the density contrast
     $\rhoi/\rhoe$, profile steepness $\mu$, and the half-width-to-length-ratio $d/L$.  
A systematic set of example solutions was offered for parameters representative 
     of active region (AR) loops. 
 
Our results are summarized as follows, where by ``long'' (``short'') we refer to
     periodicities on the order of the axial (lateral) \Alf\ time. 
Overall, all spatio-temporal patterns consistently involve the improper contribution, 
     which nonetheless attenuates with time such that the evolution at a given location is eventually dominated by the long periodicities carried by the proper contribution. 
Physically, this attenuation is attributed to the destructive interference among the 
     improper continuum. 
Specializing to the slab axis, we demonstrate that the proper contribution
     strengthens with the density contrast, but may occasionally be stronger for less steep density profiles.
We find that short periodicities can be clearly identified in the improper contribution
     only for sufficiently localized exciters.
When identifiable, these periodicities tend to agree closely with
     the oscillation frequencies expected for the discrete leaky modes (DLMs) in classic analyses, despite that the boundary conditions therein are different
     from those in our IVP. 
We demonstrate that the eigenfunction expansion approach allows the system response to be
     interpreted as the interplay between the initial exciter and some response function, the latter depending only on the equilibrium quantities (Eqs.~\eqref{eq_figRef_c1v1} and \eqref{eq_figRef_SomgVomg}). 
All qualitative features can be explained as such, for proper and improper
    contributions alike.  

Our results enable some general remarks on the seismological applicability 
    of the extensively studied DLMs.
Before anything, the examples in our Appendix~\ref{sec_App_DLM} help clarify that
    DLMs are mathematically allowed as nontrivial solutions in classic mode analyses.
Conceptually, this does not contradict the improper continuum in any example eigenspectrum
    in Fig.~\ref{fig_DispDiagram_proper}a; our eigensolutions observe different boundary
    conditions from those in classic analyses. 
It is therefore justifiable to say that 
    ``modes that do not exist cannot be excited'' \citep[][page 22]{2023JPlPh..89e9020G},
    even though we prefer to take this as meaning the inadequacy for DLMs to account for the system evolution in the entire spatio-temporal volume.
However, this inadequacy does not invalidate previous attempts that invoke the oscillation
    frequencies or even the exponential damping rates of DLMs to diagnose, say, the physical parameters of flaring loops in the context of flare QPPs \citep[e.g.,][]{2007AstL...33..706K,2012ApJ...761..134N,2015ApJ...812...22C}.
Rather, the interference among the improper continuum eigenmodes can indeed make visible
    the DLM expectations. 
It is just that whether the DLM expectations materialize depends sensitively
    on the somehow intricate interplay between the initial exciter and the equilibrium setup. 
We take this intricacy as encouraging rather than discouraging, 
    to illustrate which point we note that DLMs are the only modes that are accepted 
    to yield short periodicities in classic mode analyses. 
Suppose that some short periodicity is observed together
    with some long periodicity, which is admittedly rare but not impossible \citep[see e.g.,][for a specific observation]{2015A&A...574A..53K}. 
The simultaneous use of both periodicities then helps alleviate the nonuniqueness issue
    for inversion problems in coronal seismology 
    \citep[see][for more on this issue]{2019A&A...622A..44A}.
Now suppose that no short periodicity can be identified in some oscillatory signal
    measured with adequate temporal cadence.
Our Fig.~\ref{fig_vAxis_critLambda} then allows one to deduce the minimal lateral extent
    of the initial exciter, thereby complementing the customary practice that focuses on diagnosing the equilibrium quantities.

\begin{acknowledgements}
This research was supported by the 
    National Natural Science Foundation of China
    (12373055, 
     41974200, 
     12273019,    
     and 
     42230203).    
We gratefully acknowledge ISSI-BJ for supporting the international team
    ``Magnetohydrodynamic wavetrains as a tool for probing the solar corona'', and ISSI-Bern for supporting the international team 
    ``Magnetohydrodynamic Surface Waves at Earth's Magnetosphere and Beyond''.
\end{acknowledgements}

\bibliographystyle{aa}
\bibliography{Bib_up2date}

\begin{thebibliography}{61}
\expandafter\ifx\csname natexlab\endcsname\relax\def\natexlab#1{#1}\fi

\bibitem[{{Andries} \& {Goossens}(2007)}]{2007PhPl...14e2101A}
{Andries}, J. \& {Goossens}, M. 2007, Physics of Plasmas, 14, 052101

\bibitem[{{Arregui} \& {Goossens}(2019)}]{2019A&A...622A..44A}
{Arregui}, I. \& {Goossens}, M. 2019, \aap, 622, A44

\bibitem[{{Aschwanden} {et~al.}(2004){Aschwanden}, {Nakariakov}, \&
  {Melnikov}}]{2004ApJ...600..458A}
{Aschwanden}, M.~J., {Nakariakov}, V.~M., \& {Melnikov}, V.~F. 2004, \apj, 600,
  458

\bibitem[{{Cally}(1986)}]{1986SoPh..103..277C}
{Cally}, P.~S. 1986, \solphys, 103, 277

\bibitem[{{Cally}(1991)}]{1991JPlPh..45..453C}
{Cally}, P.~S. 1991, Journal of Plasma Physics, 45, 453

\bibitem[{{Cally}(2003)}]{2003SoPh..217...95C}
{Cally}, P.~S. 2003, \solphys, 217, 95

\bibitem[{{Cally}(2006)}]{2006SoPh..233...79C}
{Cally}, P.~S. 2006, \solphys, 233, 79

\bibitem[{{Chen} {et~al.}(2018){Chen}, {Li}, {Kumar}, {Yu}, \&
  {Shi}}]{2018ApJ...855...47C}
{Chen}, S.-X., {Li}, B., {Kumar}, S., {Yu}, H., \& {Shi}, M. 2018, \apj, 855,
  47

\bibitem[{{Chen} {et~al.}(2015){Chen}, {Li}, {Xiong}, {Yu}, \&
  {Guo}}]{2015ApJ...812...22C}
{Chen}, S.-X., {Li}, B., {Xiong}, M., {Yu}, H., \& {Guo}, M.-Z. 2015, \apj,
  812, 22

\bibitem[{{Chen} {et~al.}(2010){Chen}, {Song}, {Li}, {Xia}, {Wu}, {Fu}, \&
  {Li}}]{2010ApJ...714..644C}
{Chen}, Y., {Song}, H.~Q., {Li}, B., {et~al.} 2010, \apj, 714, 644

\bibitem[{{De Moortel} \& {Nakariakov}(2012)}]{2012RSPTA.370.3193D}
{De Moortel}, I. \& {Nakariakov}, V.~M. 2012, Philosophical Transactions of the
  Royal Society of London Series A, 370, 3193

\bibitem[{{Decraemer} {et~al.}(2020){Decraemer}, {Zhukov}, \& {Van
  Doorsselaere}}]{2020ApJ...893...78D}
{Decraemer}, B., {Zhukov}, A.~N., \& {Van Doorsselaere}, T. 2020, \apj, 893, 78

\bibitem[{{\relax DLMF}(2016)}]{NIST:DLMF}
{\relax DLMF}. 2016, {\it NIST Digital Library of Mathematical Functions},
  {http://dlmf.nist.gov/, Release 1.0.13 of 2016-09-16}, f.~W.~J. Olver, A.~B.
  {Olde Daalhuis}, D.~W. Lozier, B.~I. Schneider, R.~F. Boisvert, C.~W. Clark,
  B.~R. Miller and B.~V. Saunders, eds.

\bibitem[{{Ebrahimi} {et~al.}(2020){Ebrahimi}, {Soler}, \&
  {Karami}}]{2020ApJ...893..157E}
{Ebrahimi}, Z., {Soler}, R., \& {Karami}, K. 2020, \apj, 893, 157

\bibitem[{{Edwin} \& {Roberts}(1982)}]{1982SoPh...76..239E}
{Edwin}, P.~M. \& {Roberts}, B. 1982, \solphys, 76, 239

\bibitem[{{Edwin} \& {Roberts}(1983)}]{1983SoPh...88..179E}
{Edwin}, P.~M. \& {Roberts}, B. 1983, \solphys, 88, 179

\bibitem[{{Edwin} \& {Roberts}(1988)}]{1988A&A...192..343E}
{Edwin}, P.~M. \& {Roberts}, B. 1988, \aap, 192, 343

\bibitem[{{Feng} {et~al.}(2011){Feng}, {Chen}, {Li}, {Song}, {Kong}, {Xia}, \&
  {Feng}}]{2011SoPh..272..119F}
{Feng}, S.~W., {Chen}, Y., {Li}, B., {et~al.} 2011, \solphys, 272, 119

\bibitem[{Goedbloed {et~al.}(2019)Goedbloed, Keppens, \&
  Poedts}]{2019CUP_goedbloed_keppens_poedts}
Goedbloed, H., Keppens, R., \& Poedts, S. 2019, Magnetohydrodynamics of
  Laboratory and Astrophysical Plasmas (Cambridge University Press)

\bibitem[{{Goedbloed} {et~al.}(2023){Goedbloed}, {Keppens}, \&
  {Poedts}}]{2023JPlPh..89e9020G}
{Goedbloed}, H., {Keppens}, R., \& {Poedts}, S. 2023, Journal of Plasma
  Physics, 89, 905890520

\bibitem[{{Goedbloed}(1998)}]{1998PhPl....5.3143G}
{Goedbloed}, J.~P. 1998, Physics of Plasmas, 5, 3143

\bibitem[{{Goossens} {et~al.}(2011){Goossens}, {Erd{\'e}lyi}, \&
  {Ruderman}}]{2011SSRv..158..289G}
{Goossens}, M., {Erd{\'e}lyi}, R., \& {Ruderman}, M.~S. 2011, \ssr, 158, 289

\bibitem[{{Guo} {et~al.}(2016){Guo}, {Chen}, {Li}, {Xia}, \&
  {Yu}}]{2016SoPh..291..877G}
{Guo}, M.-Z., {Chen}, S.-X., {Li}, B., {Xia}, L.-D., \& {Yu}, H. 2016,
  \solphys, 291, 877

\bibitem[{{Jel{\'\i}nek} \& {Karlick{\'y}}(2012)}]{2012A&A...537A..46J}
{Jel{\'\i}nek}, P. \& {Karlick{\'y}}, M. 2012, \aap, 537, A46

\bibitem[{{Karlick{\'y}} {et~al.}(2013){Karlick{\'y}},
  {M{\'e}sz{\'a}rosov{\'a}}, \& {Jel{\'\i}nek}}]{2013A&A...550A...1K}
{Karlick{\'y}}, M., {M{\'e}sz{\'a}rosov{\'a}}, H., \& {Jel{\'\i}nek}, P. 2013,
  \aap, 550, A1

\bibitem[{{Kolotkov} {et~al.}(2023){Kolotkov}, {Li}, \&
  {Leibacher}}]{2023SoPh..298...40K}
{Kolotkov}, D.~Y., {Li}, B., \& {Leibacher}, J. 2023, \solphys, 298, 40

\bibitem[{{Kolotkov} {et~al.}(2015){Kolotkov}, {Nakariakov}, {Kupriyanova},
  {Ratcliffe}, \& {Shibasaki}}]{2015A&A...574A..53K}
{Kolotkov}, D.~Y., {Nakariakov}, V.~M., {Kupriyanova}, E.~G., {Ratcliffe}, H.,
  \& {Shibasaki}, K. 2015, \aap, 574, A53

\bibitem[{{Kopylova} {et~al.}(2007){Kopylova}, {Melnikov}, {Stepanov}, {Tsap},
  \& {Goldvarg}}]{2007AstL...33..706K}
{Kopylova}, Y.~G., {Melnikov}, A.~V., {Stepanov}, A.~V., {Tsap}, Y.~T., \&
  {Goldvarg}, T.~B. 2007, Astronomy Letters, 33, 706

\bibitem[{{Li} {et~al.}(2020){Li}, {Antolin}, {Guo}, {Kuznetsov}, {Pascoe},
  {Van Doorsselaere}, \& {Vasheghani Farahani}}]{2020SSRv..216..136L}
{Li}, B., {Antolin}, P., {Guo}, M.~Z., {et~al.} 2020, \ssr, 216, 136

\bibitem[{{Li} {et~al.}(2022){Li}, {Chen}, \& {Li}}]{2022ApJ...928...33L}
{Li}, B., {Chen}, S.-X., \& {Li}, A.-L. 2022, \apj, 928, 33

\bibitem[{{Li} {et~al.}(2018){Li}, {Guo}, {Yu}, \&
  {Chen}}]{2018ApJ...855...53L}
{Li}, B., {Guo}, M.-Z., {Yu}, H., \& {Chen}, S.-X. 2018, \apj, 855, 53

\bibitem[{{Lim} {et~al.}(2020){Lim}, {Nakariakov}, {Yu}, {Cho}, \&
  {Moon}}]{2020ApJ...893...62L}
{Lim}, D., {Nakariakov}, V.~M., {Yu}, D.~J., {Cho}, I.-H., \& {Moon}, Y.-J.
  2020, \apj, 893, 62

\bibitem[{{Meerson} {et~al.}(1978){Meerson}, {Sasorov}, \&
  {Stepanov}}]{1978SoPh...58..165M}
{Meerson}, B.~I., {Sasorov}, P.~V., \& {Stepanov}, A.~V. 1978, \solphys, 58,
  165

\bibitem[{{Nakariakov} {et~al.}(2021){Nakariakov}, {Anfinogentov}, {Antolin},
  {Jain}, {Kolotkov}, {Kupriyanova}, {Li}, {Magyar}, {Nistic{\`o}}, {Pascoe},
  {Srivastava}, {Terradas}, {Vasheghani Farahani}, {Verth}, {Yuan}, \&
  {Zimovets}}]{2021SSRv..217...73N}
{Nakariakov}, V.~M., {Anfinogentov}, S.~A., {Antolin}, P., {et~al.} 2021, \ssr,
  217, 73

\bibitem[{{Nakariakov} {et~al.}(2022){Nakariakov}, {Banerjee}, {Li}, {Wang},
  {Zimovets}, \& {Falanga}}]{2022SSRv..218...13N}
{Nakariakov}, V.~M., {Banerjee}, D., {Li}, B., {et~al.} 2022, \ssr, 218, 13

\bibitem[{{Nakariakov} {et~al.}(2012){Nakariakov}, {Hornsey}, \&
  {Melnikov}}]{2012ApJ...761..134N}
{Nakariakov}, V.~M., {Hornsey}, C., \& {Melnikov}, V.~F. 2012, \apj, 761, 134

\bibitem[{{Nakariakov} \& {Kolotkov}(2020)}]{2020ARA&A..58..441N}
{Nakariakov}, V.~M. \& {Kolotkov}, D.~Y. 2020, \araa, 58, 441

\bibitem[{{Nakariakov} \& {Ofman}(2001)}]{2001A&A...372L..53N}
{Nakariakov}, V.~M. \& {Ofman}, L. 2001, \aap, 372, L53

\bibitem[{{Nakariakov} {et~al.}(1999){Nakariakov}, {Ofman}, {Deluca},
  {Roberts}, \& {Davila}}]{1999Sci...285..862N}
{Nakariakov}, V.~M., {Ofman}, L., {Deluca}, E.~E., {Roberts}, B., \& {Davila},
  J.~M. 1999, Science, 285, 862

\bibitem[{{Oliver} {et~al.}(2014){Oliver}, {Ruderman}, \&
  {Terradas}}]{2014ApJ...789...48O}
{Oliver}, R., {Ruderman}, M.~S., \& {Terradas}, J. 2014, \apj, 789, 48

\bibitem[{{Oliver} {et~al.}(2015){Oliver}, {Ruderman}, \&
  {Terradas}}]{2015ApJ...806...56O}
{Oliver}, R., {Ruderman}, M.~S., \& {Terradas}, J. 2015, \apj, 806, 56

\bibitem[{Richtmyer(1978)}]{richtmyer1978principles}
Richtmyer, R. 1978, Principles of Advanced Mathematical Physics, Principles of
  Advanced Mathematical Physics No. v. 1 (Springer Verlag)

\bibitem[{Roberts(2019)}]{2019CUP_Roberts}
Roberts, B. 2019, MHD Waves in the Solar Atmosphere (Cambridge University
  Press)

\bibitem[{{Ruderman} \& {Roberts}(2006{\natexlab{a}})}]{2006SoPh..237..119R}
{Ruderman}, M.~S. \& {Roberts}, B. 2006{\natexlab{a}}, \solphys, 237, 119

\bibitem[{{Ruderman} \& {Roberts}(2006{\natexlab{b}})}]{2006JPlPh..72..285R}
{Ruderman}, M.~S. \& {Roberts}, B. 2006{\natexlab{b}}, Journal of Plasma
  Physics, 72, 285

\bibitem[{{Schrijver}(2007)}]{2007ApJ...662L.119S}
{Schrijver}, C.~J. 2007, \apjl, 662, L119

\bibitem[{{Shi} {et~al.}(2023){Shi}, {Li}, {Chen}, {Guo}, \&
  {Yuan}}]{2023ApJ...943L..19S}
{Shi}, M., {Li}, B., {Chen}, S.-X., {Guo}, M., \& {Yuan}, S. 2023, \apjl, 943,
  L19

\bibitem[{{Shibata} \& {Magara}(2011)}]{2011LRSP....8....6S}
{Shibata}, K. \& {Magara}, T. 2011, Living Reviews in Solar Physics, 8, 6

\bibitem[{{Snyder} \& {Love}(1983)}]{1983optical..book....S}
{Snyder}, A.~W. \& {Love}, J. 1983, {Optical Waveguide Theory} (Springer)

\bibitem[{{Soler} \& {Terradas}(2015)}]{2015ApJ...803...43S}
{Soler}, R. \& {Terradas}, J. 2015, \apj, 803, 43

\bibitem[{{Spruit}(1982)}]{1982SoPh...75....3S}
{Spruit}, H.~C. 1982, \solphys, 75, 3

\bibitem[{{Terradas} {et~al.}(2007){Terradas}, {Andries}, \&
  {Goossens}}]{2007SoPh..246..231T}
{Terradas}, J., {Andries}, J., \& {Goossens}, M. 2007, \solphys, 246, 231

\bibitem[{{Terradas} {et~al.}(2005){Terradas}, {Oliver}, \&
  {Ballester}}]{2005A&A...441..371T}
{Terradas}, J., {Oliver}, R., \& {Ballester}, J.~L. 2005, \aap, 441, 371

\bibitem[{{Van Doorsselaere} {et~al.}(2016){Van Doorsselaere}, {Kupriyanova},
  \& {Yuan}}]{2016SoPh..291.3143V}
{Van Doorsselaere}, T., {Kupriyanova}, E.~G., \& {Yuan}, D. 2016, \solphys,
  291, 3143

\bibitem[{{Verwichte} {et~al.}(2005){Verwichte}, {Nakariakov}, \&
  {Cooper}}]{2005A&A...430L..65V}
{Verwichte}, E., {Nakariakov}, V.~M., \& {Cooper}, F.~C. 2005, \aap, 430, L65

\bibitem[{{Wang} {et~al.}(2023){Wang}, {Li}, {Chen}, \&
  {Shi}}]{2023ApJ...943...91W}
{Wang}, Z., {Li}, B., {Chen}, S.-X., \& {Shi}, M. 2023, \apj, 943, 91

\bibitem[{Whitham(1974)}]{1974Whitham}
Whitham, G. 1974, Linear and Nonlinear Waves (Wiley)

\bibitem[{{Yu} {et~al.}(2015){Yu}, {Li}, {Chen}, \&
  {Guo}}]{2015ApJ...814...60Y}
{Yu}, H., {Li}, B., {Chen}, S.-X., \& {Guo}, M.-Z. 2015, \apj, 814, 60

\bibitem[{{Yu} {et~al.}(2016){Yu}, {Nakariakov}, \&
  {Yan}}]{2016ApJ...826...78Y}
{Yu}, S., {Nakariakov}, V.~M., \& {Yan}, Y. 2016, \apj, 826, 78

\bibitem[{{Zajtsev} \& {Stepanov}(1975)}]{1975IGAFS..37....3Z}
{Zajtsev}, V.~V. \& {Stepanov}, A.~V. 1975, Issledovaniia Geomagnetizmu
  Aeronomii i Fizike Solntsa, 37, 3

\bibitem[{{Zimovets} {et~al.}(2021){Zimovets}, {McLaughlin}, {Srivastava},
  {Kolotkov}, {Kuznetsov}, {Kupriyanova}, {Cho}, {Inglis}, {Reale}, {Pascoe},
  {Tian}, {Yuan}, {Li}, \& {Zhang}}]{2021SSRv..217...66Z}
{Zimovets}, I.~V., {McLaughlin}, J.~A., {Srivastava}, A.~K., {et~al.} 2021,
  \ssr, 217, 66

\end{thebibliography}

\begin{appendix}

\section{Abbreviations}
\label{sec_App_abbr}
This appendix lists the abbreviations in alphabetical order for the ease of reference. \\

\begin{tabular}{ll}
AR     & active region \\
BC     & boundary condition \\
BVP    & boundary value problem\\ 
DLM    & discrete leaky mode\\
DR     & dispersion relation\\
EUV    & Extreme ultraviolet\\
EVP    & eigenvalue problem\\
IBVP   & initial boundary value problem\\
IC     & initial condition\\
IVP    & initial value problem\\
MHD    & magnetohydrodynamics/magnetohydrodynamic\\
ODE    & ordinary differential equation\\
PFLK   & principal fast leaky kink mode\\
QPP    & quasi-periodic pulsation \\
RHS    & right-hand side\\
\end{tabular}

\section{Discrete leaky modes (DLMs)}
\label{sec_App_DLM}
This appendix provides some necessary details for the discrete leaky modes (DLMs),
     complementing our examination on kink motions in the main text.
 
Consider an equilibrium that differs from the main text only in that
     the domain is open in the uniform $z$-direction as well ($-\infty < z < \infty$).
Restrict ourselves to strictly 2D kink perturbations.
We Fourier-decompose any perturbation $f_1(x,z,t)$ as 
     $\tilde{f}(x)\exp\left[-i(\Omega t-kz)\right]$,
     with $\tilde{f}$ being the Fourier amplitude.
A ``classic BVP'' then results
     from linear, pressureless, ideal MHD.
Specifically, nontrivial solutions are sought for 
    \begin{equation}
        -\vA^2(x)
             \left( 
    	          \dfrac{\mathd^2}{\mathd x^2}\tilde{v} 
     			      - k^2 \tilde{v}
              \right)
    = \Omega^2 \tilde{v}, 
    \label{eq_app_BVP_govE}
    \end{equation}
   defined on $[0, \infty)$ and subject to the BCs
    \begin{subequations}
    \label{eq_app_BVP_BC}
    \begin{align}
       & \mathd\tilde{v}/\mathd x (x=0) = 0, \\
       & \mbox{no ingoing waves at $x\to\infty$}.
    \end{align}
    \end{subequations} 
This classic BVP is almost identical to the EVP in 
    Sect.~\ref{sec_sub_SolOverall}, the only difference being the BC at infinity (Eq.~(\ref{eq_app_BVP_BC}b)).

The notations in what follows are understood to apply only in this appendix.
We start by introducing the definitions,
    \begin{subequations}
    \begin{align}
    &  \nui^2 \coloneqq \dfrac{\Omega^2 - k^2\vAi^2}{\vAi^2} 
                      = \dfrac{\Omega^2}{\vAi^2}-k^2, \\
    &  \nue^2 \coloneqq \dfrac{\Omega^2 - k^2\vAe^2}{\vAe^2} 
                      = \dfrac{\Omega^2}{\vAe^2}-k^2, \\
    &  D \coloneqq      \dfrac{\Omega^2}{\vAi^2}-\dfrac{\Omega^2}{\vAe^2},  
    \end{align}
    \label{eq_def_muimue}%
    \end{subequations}     
    together with their dimensionless counterparts
    \begin{equation}
    \label{eq_def_barmuimue}
        \barnui  \coloneqq \nui d,     \quad  
        \barnue  \coloneqq \nue d,     \quad  
        \barD    \coloneqq D d^2.  
    \end{equation}
We refer to $\Re\Omega$ as the oscillation frequency, 
   seeing it as positive without loss of generality.
We take $-\pi<\arg \barD\le \pi, -\pi/2<\arg \barnui, \arg \barnue \le \pi/2$. 
The solution in the uniform exterior always writes $\propto \Exp{i\nue x}$
    to observe Eq.~(\ref{eq_app_BVP_BC}b). 
This BC turns out to be crucial in that the concept
    of dispersion relations (DRs) always makes sense, one shortcut derivation being to replace $\kappae$ with $-i\nue$ in the DRs for proper eigenmodes in the main text. 
The set $\{\Omega\}$, referred to as the spectrum of the classic BVP, 
    comprises an infinity of discrete values for any axial wavenumber
    $kd$ given a combination $[\rhoi/\rhoe, \mu]$.
By ``mode'' we refer to a nontrivial solution, 
    and we accordingly call $\Omega$ a mode frequency.

\begin{figure}
\centering
\includegraphics[width=.95\columnwidth]{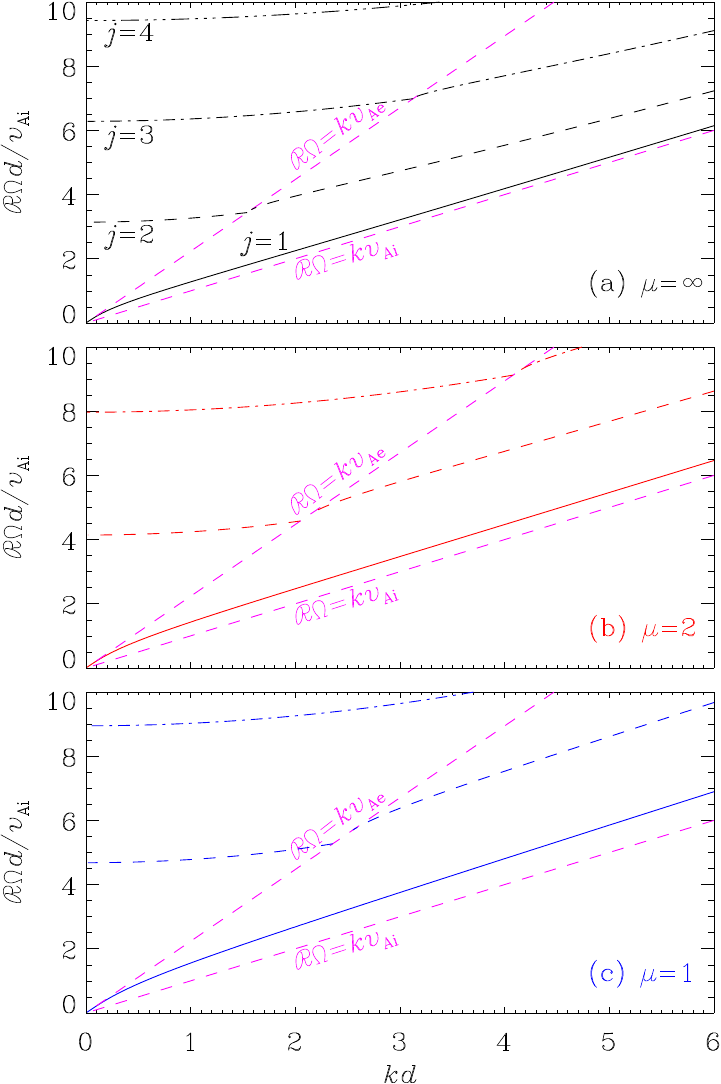}
\caption{Dependence on the axial wavenumber ($k$) of
    the real part of the mode frequency ($\Re\Omega$) 
    pertaining to the classic BVP. 
The density contrast is fixed at $\rhoi/\rhoe=5$, whereas several
    steepness parameters ($\mu$) are examined in different panels as labeled. 
The magenta dashed lines represent $\Re\Omega = k\vAi$ and $\Re\Omega=k\vAe$, 
    the latter separating
    discrete leaky modes (DLMs, to its left) from trapped modes (to its right). 
The trapped modes are identical to the proper eigenmodes in
    Fig.~\ref{fig_DispDiagram_proper}; the mode labeling convention
    therein applies here.
Mode labels with $j\ge 2$ also make sense for the DLMs, as illustrated in panel a.  
}
\label{fig_app_DispDiagram_classic} 
\end{figure}

We choose to write down the DRs for the BVP explicitly.
Now recall that the equilibrium density 
   is prescribed by Eqs.~\eqref{eq_prof_gen} and \eqref{eq_prof_innermu}.
For step profile (i.e., $\mu=\infty$), the DR writes 
    \begin{eqnarray}
        \nui\tan(\nui d) = -i\nue.
       \label{eq_step_classicDR}
    \end{eqnarray} 
For $\mu=2$, the DR reads
\begin{equation}
      i\barnue 
    = -p + 4p\alpha \dfrac{M(\alpha+1, 3/2, p)}{M(\alpha, 1/2, p)},
    \label{eq_mu02_classicDR}
\end{equation}
    where $\alpha$ and $p$ are defined by
    \begin{equation*}
             p \coloneqq \sqrt{\barD}, \quad
        \alpha \coloneqq \dfrac{1}{4}-\dfrac{{\barnui}^2}{4p}.
    \end{equation*}
The DR for $\mu=1$ writes
    \begin{equation}
        \dfrac{\Ai'(X_1) \Bi'(X_0)  -\Ai'(X_0) \Bi'(X_1)}
              {\Ai( X_1)  \Bi'(X_0) -\Ai'(X_0) \Bi( X_1)} 
    =   \dfrac{i\barnue}{\bar{D}^{1/3}},
        \label{eq_mu01_classicDR}
    \end{equation}
    where
    \begin{equation*}
        X_0 = \dfrac{-\barnui^2}{\barD^{2/3}}, \quad
        X_1 = \dfrac{-\barnui^2+\barD}{\barD^{2/3}}.
    \end{equation*}
To our knowledge, Eq.~\eqref{eq_step_classicDR} was first given 
    by \citet{2005A&A...441..371T}.
The DRs for the two continuous density profiles, however, are not available
    in the literature.

Figure~\ref{fig_app_DispDiagram_classic} fixes the density contrast
    at $\rhoi/\rhoe=5$ to illustrate the general mode behavior.
Displayed here are the $k$-dependencies of the oscillatory
    frequencies ($\Re\Omega$) for a number of steepness parameters
    ($\mu$) in different panels as labeled. 
The two magenta dashed lines in each panel represent $\Re\Omega = k\vAi$ and 
    $\Re\Omega=k\vAe$. 
All panels are qualitatively the same regarding the mode behavior, and hence
    it suffices to consider only Fig.~\ref{fig_app_DispDiagram_classic}a where the simplest step case ($\mu=\infty$) is examined.
Modes are allowed only when $\Re\Omega > k\vAi$, 
    and the spectrum at any $k$ comprises two subsets. 
One subset contains a finite number of real-valued $\Omega$, 
    which consistently satisfy $\Re\Omega=\Omega < k\vAe$.  
The associated modes correspond to $\arg \barnue = \pi/2$,
    and are the well known trapped modes \citep[e.g.,][Sect.~5.5]{2019CUP_Roberts}.
They are identical to the proper eigenmodes examined in the main text,
    making it possible to invoke the concept of cutoff axial wavenumbers
    ($\kcutj$ with $j\ge 2$).
The labeling convention in Fig.~\ref{fig_DispDiagram_proper} 
   therefore holds here.
An infinite number of complex-valued $\Omega$ are present in another subset.
These are what we call DLMs, a defining feature being $\Im\Omega <0$.
We label a DLM in view of its connection to the pertinent trapped branch.    As noted in \citet{2005A&A...441..371T}, 
    something subtle happens when $k$ is only marginally smaller than 
    some $\kcutj$. 
Take $j=2$ for instance, as shown by the dashed curve
    in Fig.~\ref{fig_app_DispDiagram_classic}a.        
By subtle we refer to the fact that no DLM exists in a narrow interval
    immediately
    on the left of $k_{{\rm cut}, 2}$, meaning some broken connection
    between the DLM portion and the trapped portion.
That said, this connection remains straightforward to identify for $j=2$,
    as is also the case for $j\ge 3$.
Now recall that all computations in the main text pertain to
    a $kd$ that is consistently smaller than $k_{{\rm cut}, 2}$.
The pertinent set of DLMs therefore starts with $j=2$. 
\end{appendix}

\end{document}